\documentclass[sigplan,nonacm]{acmart}
\AtBeginDocument{%
  }

\usepackage{tikz}
\usepackage{amsmath}
\usepackage[linesnumbered,ruled,vlined]{algorithm2e}
\usepackage{subcaption}
\usepackage{xspace}
\usepackage{booktabs}
\usepackage{amsthm} 
\usepackage{multirow}
\usepackage{paralist}

\newcommand{\SysName}{SLOs-Serve\xspace}
\newcommand{\Todo}[1]{}
\newcommand{\ZJ}[1]{}
\newcommand{\Siyuan}[1]{}
\newcommand{\Phil}[1]{}
\SetKwInput{KwState}{State}

\renewcommand{\keywords}[1]{}

\renewcommand\footnotetextcopyrightpermission[1]{}
\begin{document}

\title{\SysName: 
Optimized Serving of Multi-SLO LLMs}

\author{Siyuan Chen}
\authornote{Part of this work was done during an internship at Google.}
\email{siyuanc3@andrew.cmu.edu}
\affiliation{
    \institution{Carnegie Mellon University}
    \city{Pittsburgh}
    \state{PA}
    \country{USA}
}
\author{Zhipeng Jia}
\email{zhipengjia@google.com}
\affiliation{
    \institution{Google}
    \city{Seattle}
    \state{WA}
    \country{USA}
}
\author{Samira Khan}
\email{samirakhan@google.com}
\affiliation{
    \institution{Google}
    \city{New York}
    \state{NY}
    \country{USA}
}
\author{Arvind Krishnamurthy}
\email{arvindkrish@google.com}
\affiliation{
    \institution{Google}
    \city{Seattle}
    \state{WA}
    \country{USA}
}
\author{Phillip B. Gibbons}
\email{pgibbons@andrew.cmu.edu}
\affiliation{
    \institution{Carnegie Mellon University}
    \city{Pittsburgh}
    \state{PA}
    \country{USA}
}

\begin{abstract}
This paper introduces \SysName, a system designed for serving multi-stage large language model (LLM) requests with application- and stage-specific service level objectives (SLOs). 
The key idea behind \SysName is to customize the allocation of tokens to meet these SLO requirements.
\SysName uses a multi-SLO dynamic programming-based algorithm to continuously optimize token allocations under SLO constraints by exploring the full design space of chunked prefill and (optional) speculative decoding. 
Leveraging this resource planning algorithm, \SysName effectively supports multi-SLOs and multi-replica serving with dynamic request routing while being resilient to bursty arrivals. 
Our evaluation across 6 LLM application scenarios (including summarization, coding, chatbot, tool calling, and reasoning) demonstrates that \SysName improves per-GPU serving capacity by 2.2x on average compared to prior state-of-the-art systems.
\end{abstract}



\keywords{Large Language Models, Service Level Objectives, Serving System, Cloud Computing}


\maketitle

\section{Introduction}
Large Language Models (LLMs) are becoming increasingly popular. 
Current LLM applications, such as interactive chatbot~\cite{openai_chatgpt}, coding assistant~\cite{microsoft_copilot}, document summarizer~\cite{openai_summary}, often feature \textit{multi-stage} processing: A \textit{prefill} stage processes the input text, and a \textit{decode} stage generates tokens one by one. More recent \textit{Reasoning LLMs}~\cite{guo2025deepseek} introduce a new thinking stage between prefill and decode for solving highly complex problems. And, when moving to agentic scenarios where LLMs are given access to a list of tools to solve real-world problems~\cite{qin2023toolllm}, a request is returned after rounds of internal processing stages between the LLM and the tools.


\newcommand{\stagewidth}{0.55\linewidth}
\begin{table}[t]
    \centering
    \caption{Multi-stage LLM applications with diverse SLOs.}\label{tab:stages}
\footnotesize{
    \begin{tabular}{@{}c@{}c@{}p{2.05cm}@{}}
        \toprule
        \textbf{Application} & \textbf{Stages} & \textbf{SLOs}\\
        \midrule
        Summarization & 
        \begin{minipage}[t]{\stagewidth}
            \begin{itemize}
                \item Long, fast prompt processing
                \item Short reading-speed response
            \end{itemize}
        \end{minipage}
        & Tight on prefill\newline Loose on decode\\
        \midrule
        Coder & 
        \begin{minipage}[t]{\stagewidth}
            \begin{itemize}
                \item Normal length/speed prompt
                \item Long, fast response
            \end{itemize}
        \end{minipage}
        & Loose on prefill\newline Tight on decode\\
        \midrule
        ChatBot & 
        \begin{minipage}[t]{\stagewidth}
            \begin{itemize}
                \item Normal length/speed prompt
                \item Reading-speed response
            \end{itemize}
        \end{minipage}
        & Loose on prefill\newline Loose on decode\\
        \midrule
        LLM with Tools & 
        \begin{minipage}[t]{\stagewidth}
            \begin{itemize}
                \item Normal length/speed prompt
                \item Fast toolCall-toolResponse loop
                \item Reading-speed response
            \end{itemize}
        \end{minipage}
        & Tight on prefill\newline Tight on pref./dec.\newline Loose on decode\\
        \midrule
        Reasoning LLM & 
        \begin{minipage}[t]{\stagewidth}
            \begin{itemize}
                \item Normal length/speed prompt
                \item Long fast thinking
                \item Reading-speed response
            \end{itemize}
        \end{minipage}
        & Tight on prefill \newline Tight on decode\newline Loose on decode\\
        \bottomrule
    \end{tabular}
   }
\end{table}
\begin{figure}[t]
    \includegraphics{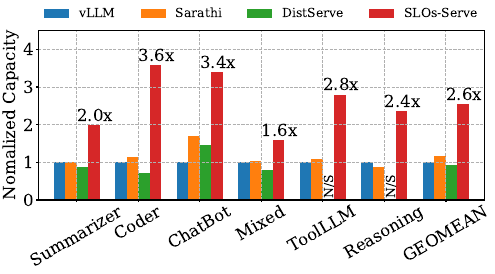}
    \caption{Serving Capacity comparison for LLM applications with heterogeneous SLOs, on a server with 4 A100s (experimental details in Tables~\ref{tab:scenario} and~\ref{tab:datasets} of \S\ref{sec:eval}). N/S: not supported.}
    \label{fig:teaser}
\end{figure}
An LLM serving system, hosted on shared cloud servers, delivers online responses to LLM requests under service level objectives (SLOs).
Meeting SLOs plays a key role in ensuring a positive user experience. As summarized in Table~\ref{tab:stages}, different stages in LLM serving often demand stage-specific SLOs for the best user experience. 
For example, in prefill-decode-based applications, document summarization demands high prefill throughput to digest lengthy documents, chatbot demands smooth decoding flow to match a human's reading speed, and coding assistant demands lower per-token decode latency for code generation. What's more, to respond to the user sooner, reasoning models may want to squeeze the time for the thinking stage, and an LLM-with-tools application may want to minimize the latency back and forth between the LLM and the tools. 


Supporting application-specific SLOs across multi-stage processings presents unique challenges in LLM serving due to complex resource sharing across requests and stages. State-of-the-art (SOTA) serving systems~\cite{vllm,sarathi,zhong2024distserve,patel2024splitwise,sun2024llumnix,hu2024inference} adopt continuous batching~\cite{yu2022orca} to maximize the serving throughput by forming batches comprising of \textit{tokens} from different stages and requests.
Moreover, LLM serving systems use either co-located or disaggregated approaches to map processing stages to GPUs. Both approaches exhibit unique challenges for the scheduler to meet per-stage SLOs (\S\ref{sec:background_challenge}). Co-location runs all stages on the same GPUs, with schedulers like vLLM~\cite{vllm} and Sarathi-Serve~\cite{sarathi} prioritizing different stages in isolation. This greedy approach causes significant SLO violations, especially during traffic spikes, as illustrated later in Fig.~\ref{fig:comparison}. 
The disaggregation approach~\cite{patel2024splitwise,zhong2024distserve,qin2024mooncake,hu2024memserve,hu2024inference}, on the other hand, separates stages onto different devices with customized configurations. But the approach struggles when load distribution changes dynamically due to varying input/output lengths or shifts in application domains (as illustrated later in Fig.~\ref{fig:profile-disagg}).

In this paper, we propose \textit{\SysName}, an LLM serving system designed to handle LLM applications with per-stage or per-request SLOs, overcoming all the above challenges. \SysName introduces an SLO-optimized scheduling algorithm tailored for LLM requests (\S\ref{sec:sch-alg}). The key idea is to customize token allocation to meet stage- and request-specific SLO requirements. 
Unlike prior schedulers that greedily attain the SLO of a single stage, \SysName's scheduler identifies a set of requests with attainable SLOs and, for that set, guarantees to satisfy every stage's SLO until completion.
Specifically, \SysName designs a multi-SLO dynamic programming-based algorithm (\S\ref{sec:dp-alg}) to continuously optimize token allocations in batches under stage- and request-dependent SLO constraints by exploring the full design space of chunked prefill and dynamic batch-size tuning (\S\ref{sec:bs-tuning}).  
Additionally, we propose \textit{SLO-adaptive speculative decoding} (\S\ref{sec:slo-ada-sd}), which customizes the speculation length in speculative decoding to meet stage- and request-dependent SLOs. 
The accuracy of our optimization depends on a profiling-based characterization of the target GPU's capabilities on LLM workloads, which varies across GPU families (e.g., A100s vs.~H100s).

Building on top of the scheduling algorithm, we prototype \SysName, a distributed LLM serving system that supports multi-SLOs and multi-replica serving, while being resilient to bursty arrivals. 
During temporary high request loads, \SysName delays requests with unattainable SLOs to secure the SLO attainment of the rest (\S\ref{sec:burst}), whereas existing schedulers' greedy approaches results in a cascading effect where every request fails to meet its SLO.
In multi-replica serving, our system runs a centralized scheduler that virtualizes every replica, and proactively routes requests with unattainable SLOs at a replica to another with minimal scheduling overhead (\S\ref{sec:multi-replica}). In this way, \SysName implements a form of \textit{soft admission control}, in which ``admitted'' requests are guaranteed to meet their multi-stage SLOs and other requests are serviced either best effort (in single GPU settings) or routed to a different GPU for servicing (in multi-GPU settings).

In evaluation, we benchmark \SysName against state-of-the-art (SOTA) serving systems (vLLM, Sarathi-Serve, DistServe) on 6 LLM application scenarios with customized SLOs. As displayed in Fig.~\ref{fig:teaser}, \SysName improves the serving capacity (defined as the maximum request load per GPU while maintaining 90\% SLO attainment) by a geo-mean of 2.2x compared to the best of Sarathi-Serve and vLLM and 2.4x compared to DistServe.
In multi-replica serving, we observed that under bursty arrivals \SysName is able to serve up to 6.2x capacity with 4-replica serving compared with 1-replica serving. The linear scaling underscores \SysName scheduling algorithm's effectiveness in load balancing.

In summary, this paper makes the following contributions:
\begin{asparaitem}
    \item We identify the challenges in multi-SLO LLM serving under fine-grained resource sharing and propose a multi-SLO dynamic programming-based scheduling algorithm to resolve the challenges. Our scheduler explores the full design space of chunked prefill, speculative decoding, and dynamic batch-size tuning, and leverages soft admission control. 
    \item We build our design into \SysName, an LLM serving system that supports multi-SLOs, multi-replica serving, and is more robust to bursty arrivals than SOTA systems. 
    \item Comprehensive evaluations across 6 application scenarios, including the first comparison study on modern tool-based and reasoning LLMs, show that \SysName is able to improve SOTA serving system's capacity by 2.2x on average.
\end{asparaitem}

\section{Background and Motivation}








In this section, we first summarize the current trend of LLM applications. We then articulate scheduling challenges introduced by emerging LLM applications while identifying pitfalls of current SOTA LLM schedulers.

\subsection{LLM Applications and Serving}

\paragraph{Multi-Stage LLM Applications} 
Emerging LLM applications are evolving beyond the traditional two-stage model (prefill and decode) toward multi-stage processing. While classic applications like chatbots and summarizers follow this basic pattern, reasoning models add a thinking stage for problem-solving. Agentic applications demonstrate even greater complexity through iterative tool interaction loops. This evolution necessitates more sophisticated resource management to handle diverse execution patterns. Table~\ref{tab:stages} summarizes characteristics of five multi-stage LLM applications.

\paragraph{Multi-SLOs Serving} 
Complex LLM processing can be modularized into prefill and decode phases with distinct service level objectives (SLOs). The prefill stage (initial prompt processing, reasoning, tool interaction) is measured by Time-To-First-Token (TTFT), which impacts user wait time. The decode stage (token generation) is measured by Time-Per-Output-Token (TPOT), affecting response delivery speed.

Stage-dependent SLO enforcement improves user experience. For reasoning models, prioritizing low TPOT in the thinking stage minimizes perceived latency, while the decode stage can tolerate higher TPOT aligned with human reading speeds. Similarly, agentic applications benefit from low TTFT and low TPOT during tool interactions and reading-speed TPOT for final responses. This granular control enables tailored optimization across different stages. 

Our goal is to maximize \textit{serving capacity}, the maximum request load each GPU can process while maintaining a target SLO attainment rate (e.g., 90\% attainment).

\subsection{LLM Serving Optimizations}

\begin{figure}[ht]
    \centering
    \includegraphics{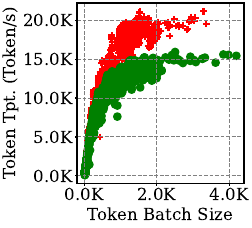}
    \includegraphics{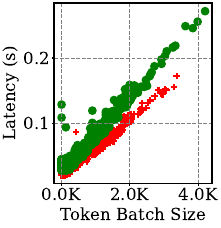}
    \caption{Throughput-latency trade-off for batching. Each data point--Green for the OPT-7B model~\cite{zhang2022opt} on A100, Red for the OPT-13B model~\cite{zhang2022opt} on H100--is a batch executed in \SysName's scheduling with both prefill and decode tokens. }
    \label{fig:batching}
\end{figure}

\textit{Continuous batching}~\cite{yu2022orca} improves throughput by combining processing stages from different requests. Since individual stages often underutilize hardware, batching increases simultaneous request handling with minimal latency overhead.
Batches contain tokens from various stages and requests—prefill contributes prompt-length tokens, while decode adds one token at a time. \textit{Token batch size} determines the throughput-latency tradeoff, with larger batches increasing throughput at the cost of latency, as shown in Fig.~\ref{fig:batching}.


Batch composition is diversified through:

\begin{asparaitem}
\item \textit{Chunked prefill}~\cite{sarathi} breaks long prefill stages into smaller chunks, preventing decode stages from stalling. Instead of using fixed maximum token sizes that limit throughput, batch sizes are dynamically adjusted based on system load (\S\ref{sec:bs-tuning}).

\item \textit{Speculative decoding}~\cite{leviathan2023fast} uses a smaller "draft" model which speculates multiple token steps. Drafted tokens are verified in a single batch by the main model. The verification step processes multiple tokens simultaneously, creating new batching opportunities. While not always beneficial due to potential errors and overhead, adapting speculation lengths to SLOs can surprisingly improve throughput (\S\ref{sec:slo-ada-sd}).
\end{asparaitem}

\subsection{Challenges}\label{sec:background_challenge}

Supporting application-specific SLOs for multi-stage processing in LLM requests presents complex challenges due to the resource contention and diversified optimizations within continuous batching.
We re-visit existing scheduling approaches and understand how they fail to meet the challenges.



\paragraph{Co-located Scheduling.}

\begin{figure}[ht]
    \centering
    \includegraphics[width=0.5\textwidth]{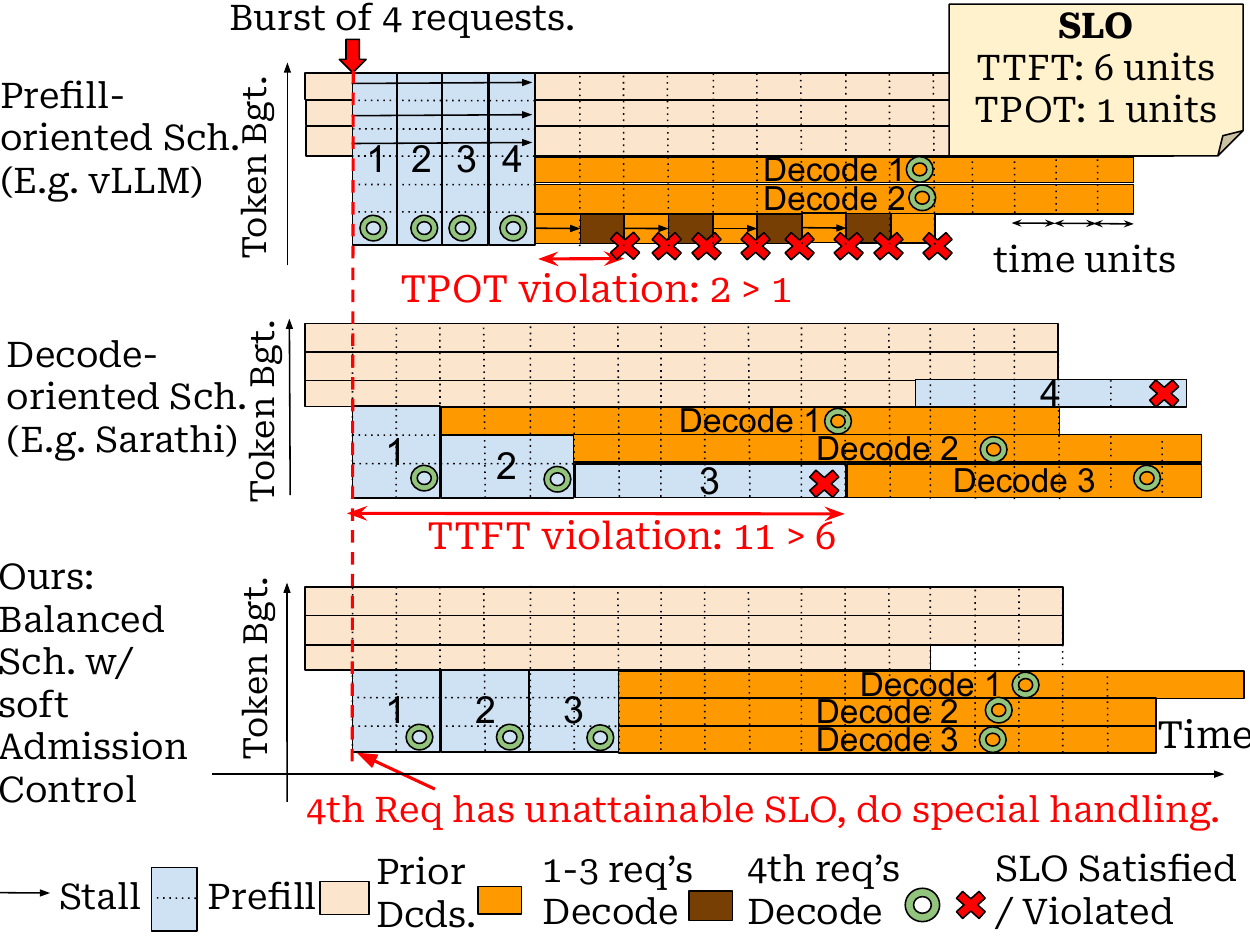}
    \caption{Comparison between different co-located scheduling approaches.}
    \label{fig:comparison}
\end{figure}

Current \textit{co-located} schedulers (all GPUs run a mix of stage types) suffer from a fundamental flaw: they prioritize attaining SLOs at individual stages in isolation. For instance, vLLM~\cite{vllm} employs a \textit{prefill-oriented scheduling} approach, which eagerly executes each request's prefill stage to minimize TTFT. But this strategy often leads to widespread TPOT violations due to the low priority assigned to decode processing. Conversely, Sarathi-serve attempts to mitigate this issue by breaking the prefill process into smaller chunks, but its \textit{decode-oriented scheduling} prioritizes filling batches with decode tokens, resulting in TTFT violations.

To illustrate these shortcomings, consider the scenario in Fig.~\ref{fig:comparison} where the system can process six tokens per time unit, and a burst of four requests arrives, each with six prefill tokens. At the time of arrival, the system has three ongoing decodes.
The SLOs are set at six time units for TTFT and one time unit for TPOT. In a prefill-oriented scheduling scheme, the scheduler preempts ongoing decoding requests to execute incoming prefill requests, causing decoding stalls and subsequent TPOT violations. As the number of decoding requests overwhelms the system's processing capacity, some requests are guaranteed to miss their TPOT SLOs. In our example, requests three and four fail to meet their TPOT requirements. In contrast, a decode-oriented scheduler prioritizes filling available resources with decode tokens, leading to prolonged prefill stages and severe TTFT violations. E.g., when the first request's prefill completes, the scheduler allocates a token budget per time unit for decoding, delaying the second request's prefill. This delay propagates, causing requests three and four to miss their TTFT SLOs, with request three's prefill completion delayed to the 11th time slot (TTFT violation) and request four's delayed even further.


The greedy nature of these scheduling approaches motivates our design of balanced scheduling with soft admission control. As shown later, our scheduler allocates tokens by taking into account the SLOs across stages (\S\ref{sec:sch-alg}) and employs special handling for requests whose SLOs are inherently unattainable (\S\ref{S:application}). In the Fig.\ref{fig:comparison} example, we successfully attain the SLOs for all existing requests, as well as three out of the four new requests.
Thus, our multi-SLO attainment rate is higher than the prior approaches.
Note that in general, some form of SLO-aware request prioritization (e.g., soft admission control) is required to prevent cascading-lateness effects and to maximize attainment rates (and serving capacity).

\paragraph{Disaggregated Scheduling}

In \textit{disaggregated} scheduling \cite{zhong2024distserve,patel2024splitwise}, different stage types are separated onto different (groups of) GPUs. In this way, one can specialize for every stage's SLO by adjusting the hardware configurations~\cite{zhong2024distserve}, and balance workloads across stages by adjusting the GPUs allocated to every stage.

\begin{figure}[h]
    \centering
    \includegraphics{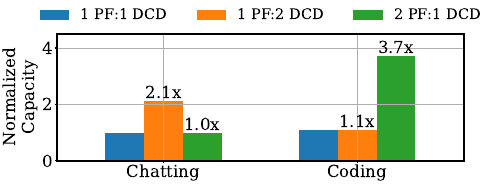}
    \caption{Capacity in DistServe with different prefill (PF), decode (DCD) device ratios when serving the OPT-13B model with H100 GPUs. The capacity is normalized to 1 PF:1 DCD.}
    \label{fig:profile-disagg}
\end{figure}


However, the flexibility of disaggregated scheduling is limited by the fact that optimal hardware configurations depend on the stage's loads, which are highly dynamic in our setup. As illustrated in Figure~\ref{fig:profile-disagg}, decode-heavy applications such as ChatBot require a higher allocation of decoding devices (1 decode: 2 prefills), whereas prefill-heavy workloads like Coder benefit from more prefill device allocation (2 prefills: 1 decode). Consequently, a single static allocation strategy is ineffective for serving multiple applications, as well as attaining request-dependent SLOs.

In fact, to achieve optimal serving capacity, the ideal prefill-to-decode device ratio in disaggregated scheduling can be expressed as:
\(\frac{n_{prefill}}{n_{decode}} =\frac{ (1-\frac{C}{TPOT})E[prefillLength]}{E[decodeLength]} \label{eqn: optimal proportion} \), 
which is dependent on both the SLO requirement and the stage loads. See Appendix~\ref{apdx:disagg-sch} for detailed discussions.

\section{\SysName Scheduler}\label{sec:sch-alg}

In this section, we introduce \SysName's scheduler that addresses the challenges discussed in \S\ref{sec:background_challenge}.

\subsection{Scheduling with Soft Admission Control}

To ensure multi-stage SLO attainment, \SysName's scheduler introduces a \textit{soft admission control} mechanism that guarantees SLO attainment for admitted requests and handles declined requests (say, 2\%) with fallback mechanisms.
Periodically, the scheduler selects an optimal subset of new requests that can attain their SLOs. 
Then, it determines execution plans, integrating these selected requests into future batched executions. 


Algorithm~\ref{alg:scheduler} shows the pseudocode of \SysName's scheduling with soft admission control. The scheduler is invoked upon the occurrence of a timeout or when the number of new or completed requests surpasses a predefined threshold. Upon invocation, the scheduler takes in states for running and new requests, as well as a performance model, to generate admission decisions and future batch schedules (Line~\ref{algln:sch}). Declined requests are (Line~\ref{algln:process-declined}) specially handled, as elaborated later in \S\ref{S:application}.
For accepted requests, the batch schedules ensure SLO attainment. Each batch is represented as:
\begin{align}
Batch := [(ID_i, S_i\in \{Prefill, Decode\}, \#Token_i)_i]\label{eqn:batch},
\end{align}
where each entry specifies the processing of $\#Token_i$ tokens for request $ID_i$ in state $S_i$. A batch supports chunked prefill by processing less than full tokens for prefill requests, as well as speculative decoding by verifying more than one token for decode requests.
The computed batches are executed using a stateless \texttt{BatchForward} function (Line~\ref{algln:exec}), which we elaborate in Algorithm~\ref{apx:batch-forward} in the Appendix.

\begin{algorithm}
\caption{\SysName's LLM Serving}\label{alg:scheduler}
\KwIn{ 
    $Reqs_{new}$
}

\KwState{
$Reqs_{running}, thresh_{new}, thresh_{finished}, \mathcal{M}:\ Perf.\ Model$
}


\textbf{Infinite Run:}\label{algln:loop}

$Reqs_{admitted}, Reqs_{declined}, Batches \gets$Schedule($Reqs_{running}$, $Reqs_{new}$, $t$, $\mathcal{M}$) \label{algln:sch}

$Reqs_{running} \gets Reqs_{running} + Reqs_{admitted}$

$Reqs_{new} \gets \emptyset$

\ForEach{request in $Reqs_{declined}$}{ \label{algln:process-declined}
    Perform best effort serving (\S\ref{sec:burst}) or route to another replica (\S\ref{sec:multi-replica}).
}

$\#finished \gets 0$

\For{$batch$ in Batches} {
    Call \texttt{BatchForward}($batch$)\; \label{algln:exec}
    \ForEach{request in $Reqs_{running}$}{
        \If{request is finished}{
            $Reqs_{running} \gets Reqs_{running} - request$
            $\#finished \gets\#finished + 1$
        }
    }
    \If {TimeOut or $\#finished > thresh_{finished}$ or $|Reqs_{new}| > thresh_{new}$} {
        Goto line~\ref{algln:loop}.
    }
}

\end{algorithm}

\subsubsection{Performance Modeling for Batch Execution.}\label{performance model}


To make scheduling decisions, the scheduler relies on a performance model that accurately characterizes per-batch execution times.
We design our model to work for different hardware architectures by adopting a generalized Roofline model~\cite{williams2009roofline}.
Specifically, the execution time of a \texttt{BatchForward} call is modeled as:
\begin{align*}
\hat{T}(\texttt{BatchForward}&(batch)) =\\
max_{l}\ (k^l_1 &\#Tokens + k^l_2 \#SpecStep + b_l)\\
\#Tokens &:= \Sigma_{i\in batch} \#Token_i\\
\#SpecStep &:= max_{i\in batch,\ S_i\ \text{is prefill}} \#Token_i 
\end{align*} where the $k^l_1, k^l_2, b_l$s are parameters obtained by regression on profiled data from runs on (multi-)GPU backends.
Each term ($l = 2$ in practice) within the max function represents a source of execution time, such as $k^l_1 \#Tokens$ for computation time, constant $b_l$ for fixed memory access to the model weights, and $k^l_2 \#SpecStep$ for the speculative model's overhead.
The max operation identifies the bottleneck that limits performance, assuming sufficient parallelism.
We validate the fidelity of the performance model against empirical measurements on A100s and H100s in Fig.~\ref{fig:perf-model}.

\subsection{\SysName's scheduling algorithm}

We now present \SysName's dynamic-programming based scheduling algorithm (Line~\ref{algln:sch} of Algorithm~\ref{alg:scheduler}).

\begin{figure}[ht]
    \centering
    \begin{subfigure}[t]{0.31\linewidth}
        \centering
\includegraphics[width=\textwidth]{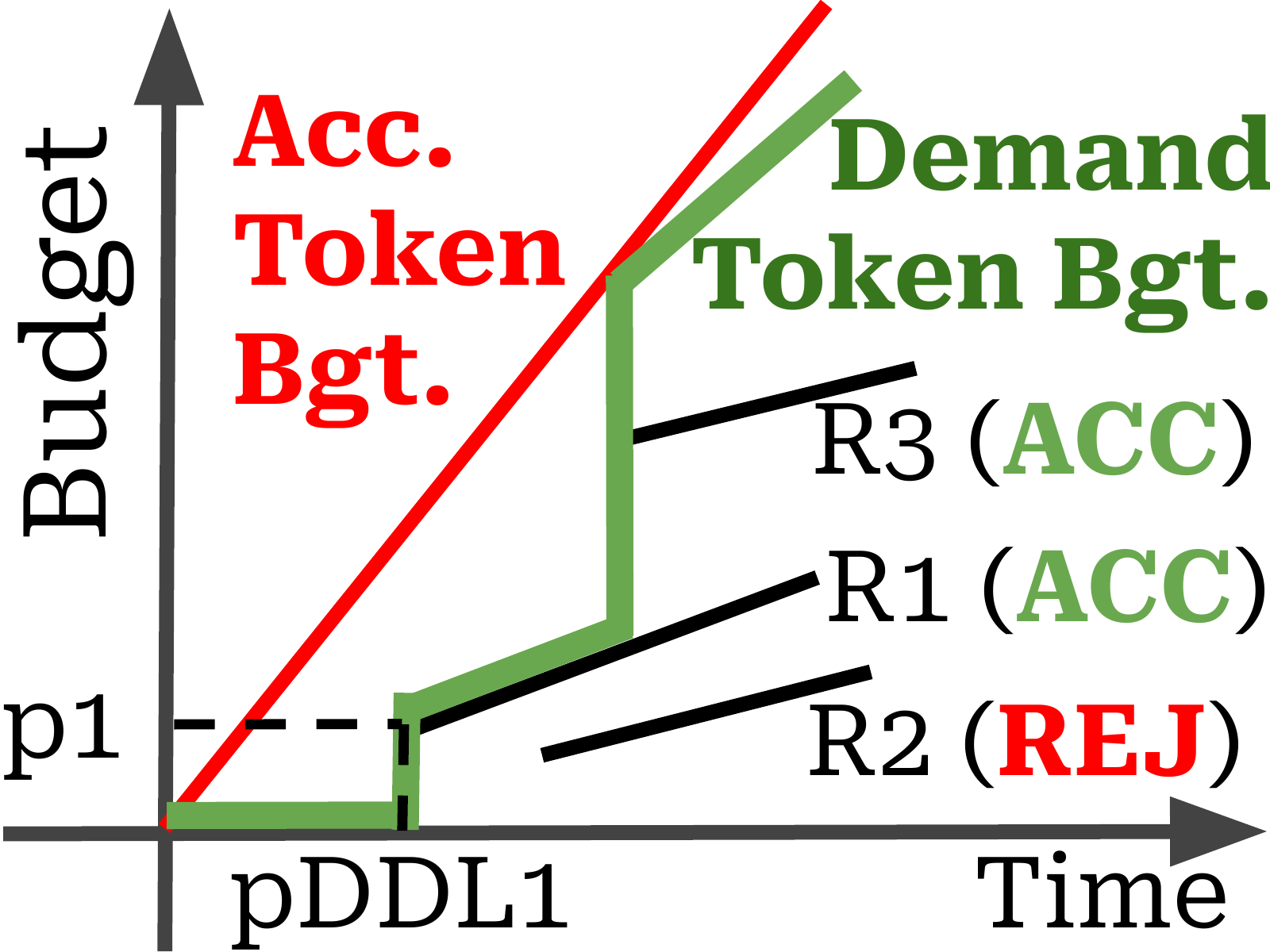}
        \caption{Valid Scheduling with admission control.}
        \label{fig:valid}
    \end{subfigure}
    \hfill
    \begin{subfigure}[t]{0.31\linewidth}
        \centering
        \includegraphics[width=\textwidth]{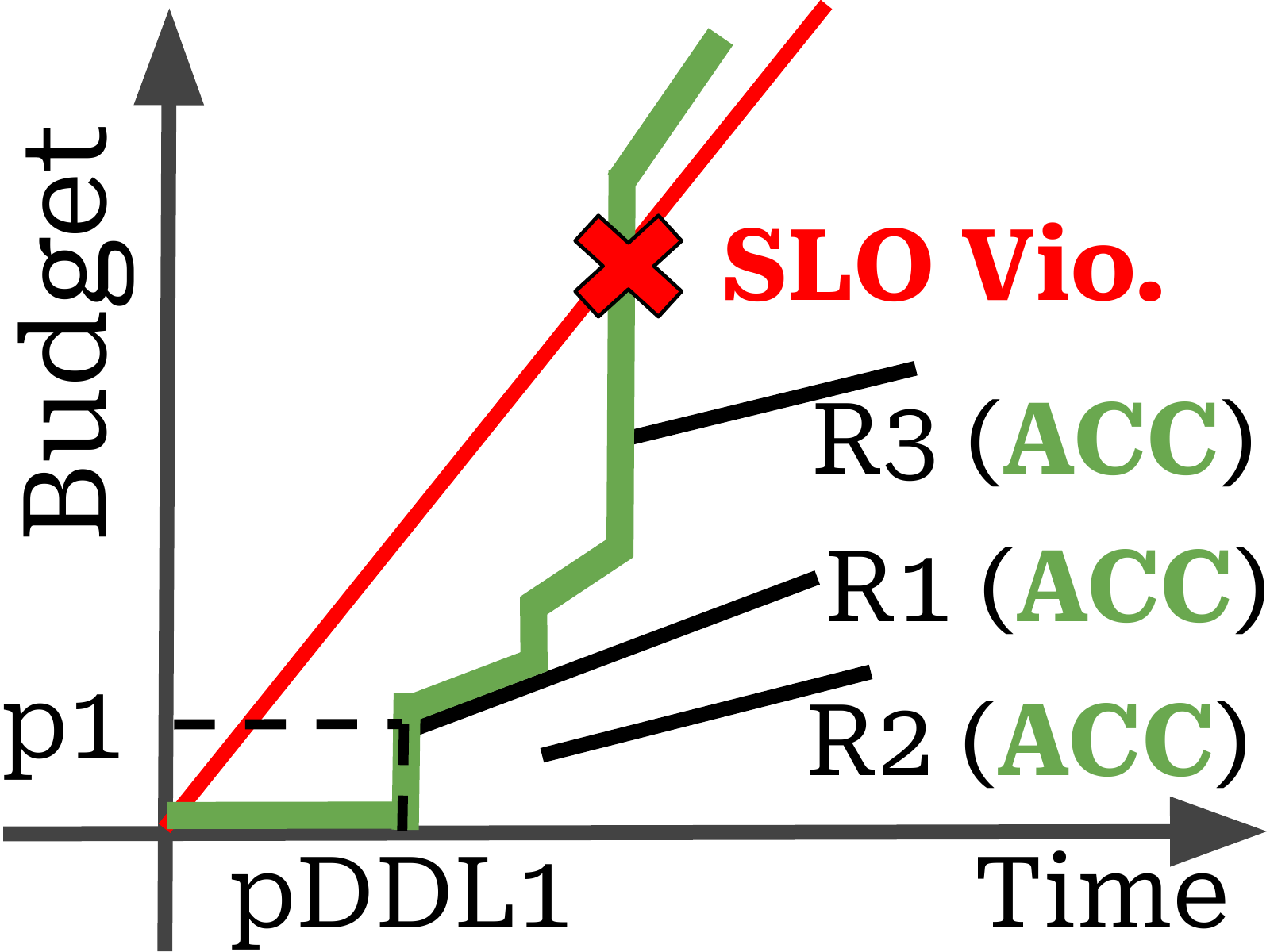}
        \caption{Invalid Scheduling w/o admission control.}
        \label{fig:invalid}
    \end{subfigure}
    \hfill
    \begin{subfigure}[t]{0.31\linewidth}
        \centering
        \includegraphics[width=\textwidth]{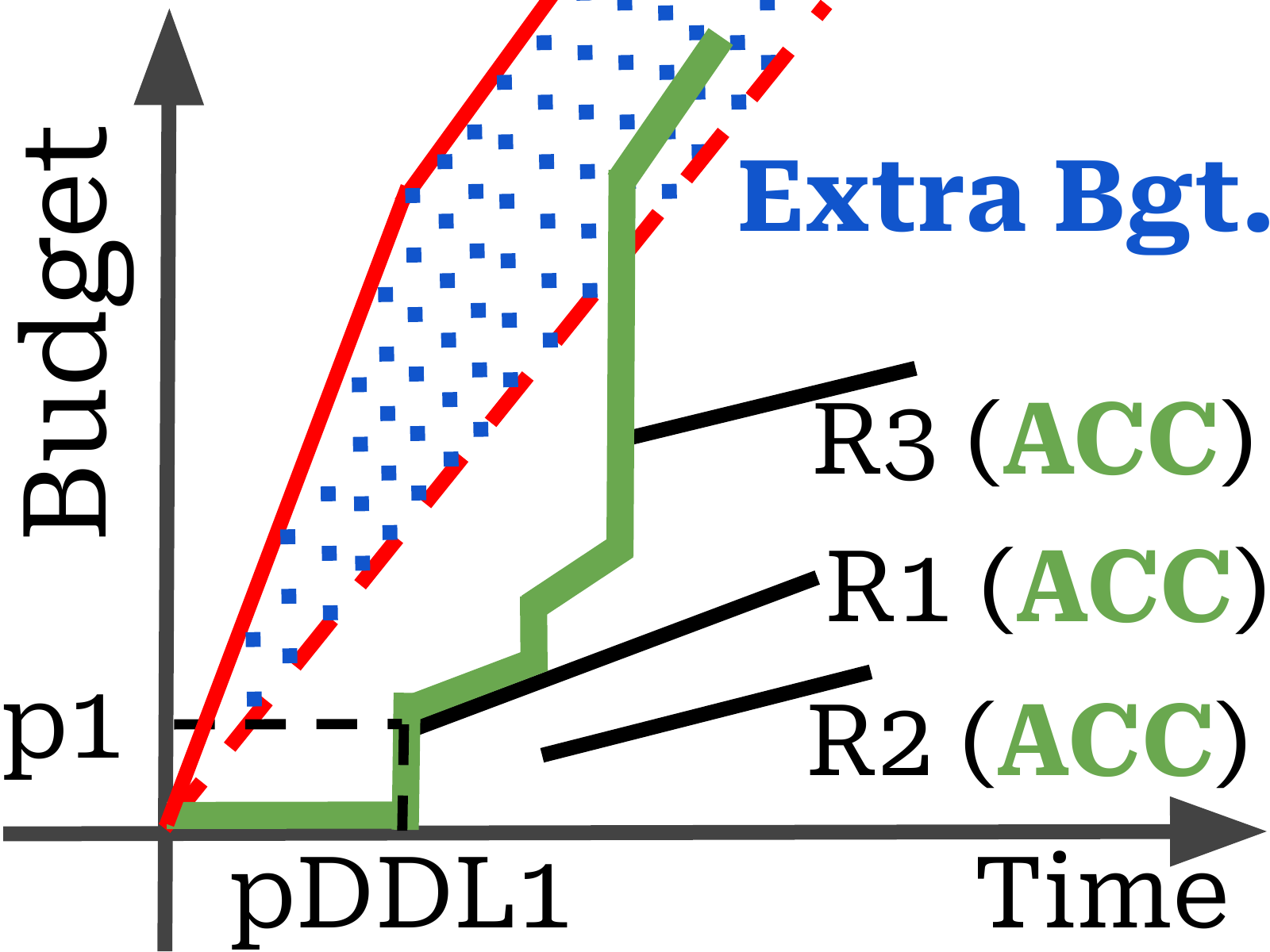}
        \caption{Gain from Dynamic Batch Size Tuning.}
        \label{fig:variance-batch}
    \end{subfigure}
    \caption{Illustration of the scheduling algorithms}
    \label{fig:illustration}
\end{figure}

\paragraph{An Example} We use a toy example in Fig.~\ref{fig:illustration} to provide the intuition. 
This example requires the scheduler to make scheduling decisions on three new requests, SLOs of which are represented by lines.
The line for request $R_i$ starts at point $(pDDL_i, p_i)$, representing that $p_i$ tokens must be allocated to $R_i$ before the prefill deadline $pDDL_i$,
and then grows at rate $k_i$ tokens/s, representing token generation with speed $k_i$. 
This representation encodes multi-SLOs across requests and stages.
For example, $R2$ may represent a chatting request with smooth decoding speed, $R1$ as a coding request with higher generation speed demand, and $R3$ as a summarization request with a longer input.

The \textit{accumulated token budget} line (red) depicts the total available tokens, the slope of which represents the token throughput decided by the launched batches.
In Fig.~\ref{fig:invalid},~\ref{fig:valid}, we assume a fixed token batch size over time, which results in a linear budget increase. 
Fig.~\ref{fig:variance-batch} demonstrates a more efficient approach through \textit{dynamic batch size tuning}, enabling non-linear budget growth with higher token throughput.

The scheduling problem---admission control and batch schedule determination---is equivalent to finding a subset of lines whose cumulative demand always remains below the accumulated token budget.
Violation of this condition, where the cumulative demand of admitted requests surpasses the budget, negates feasible schedules.
Meeting the condition, on the other hand, ensures that the system can satisfy the decoding token requirements of all admitted requests, as their combined generation rate is less than the budget's slope, and the available budget at each request's prefill deadline exceeds the accumulated prefill demand.

For example, admitting all three requests in Fig.~\ref{fig:invalid} (indicated by ACC) results in demand exceeding the budget before $R3$'s prefill deadline, leading to a violation. However, admitting only $R1$ and $R3$ (Fig.~\ref{fig:valid}) keeps the cumulative demand within the budget, indicating a valid batched schedule. 
Furthermore, implementing dynamic batch size tuning allows the attainment of SLOs for all three requests because of the enlarged token budget (Fig.~\ref{fig:variance-batch}).

\subsubsection{Scheduling via Dynamic Programming}\label{sec:dp-alg}
To make admission control and batching decisions, \SysName uses a multi-SLO dynamic programming (DP) algorithm.

We use the following notations. Suppose there are $N$ requests in total, request $R_i$ arrives at $t_i$ with prompt of length $p_i$ and a memory requirement $m_i$. $R_i$ comprises a prefill stage followed by a decode stage. Every request has a deadline for its prefill stage $pDDL_i$. For the decode stage, a request can enforce the TPOT SLO for every token from a vector of choices $TPOT_1, TPOT_2, ..., TPOT_L$. 
We say that a request's SLO is attained if and only if the SLO for every stage is satisfied. Successfully attaining $R_i$'s SLO yields value $v_i$.

The scheduling algorithm optimizes batch schedules to maximize the total value gained from attaining requests' SLO.
We solve the problem using dynamic programming that incrementally calculates $DP[i, m, pb, n]$, defined as the best value a scheduler can ever get from serving the first $i$ requests with earliest prefill deadlines, while generating $pb$ \textit{prefill budget} by the end of $R_i$'s prefill deadline ($pDDL_i$) within $m$ memory units.
The prefill budget, defined as the token budget remaining after satisfying the decode SLOs for accepted requests, is always positive and is used to attain the prefill SLOs for unscheduled requests, i.e., requests with later prefill deadlines.
The state transition function is 
\begin{align*}
&DP[i, m, pb, n] \nonumber\\
&= \max_{\substack{j,\\ pDDL_j < pDDL_i, \\ \Delta pb \ge 0,\  m \ge m_i}} \left\{ DP[j, m - m_i, pb + p_i - \Delta_{i,j}^{n-1} pb, n-1] + v_i \right\}\label{eqn:dp}.
\end{align*}
This equation enumerates over the last accepted request $j$, whose prefill deadline precedes $i$'s, to find the optimal prior state that leads to the highest value. 
To generate $pb$ prefill tokens by $pDDL_i$, the prefill budget at $pDDL_j$ must be at least $pb + p_i - \Delta_{i,j}^{n-1} pb$. Here, $p_i$ is prefill budgets consumed by $R_i$'th prefill stage, and $\Delta_{i,j}^{n-1} pb$ represents the newly generated prefill budgets during the interval between $R_i$ and $R_j$'s prefill deadlines, with $n-1$ accepted requests. Specifically, \SysName solves a constrained optimization problem to calculate the $\Delta_{i,j}^{n} pb$:
\begin{align}
&\Delta_{i,j}^{n} pb = PB^*(pDDL_i - pDDL_j, n), \text{and } \\
&PB^*(t, n) = \max_{\text{partial batches}} \sum_{b \in \text{partial batches}} b.\text{prefillBudget},\nonumber \\
&\quad \text{subject to attaining decode SLOs for } n \ \text{requests}\label{eqn:pb}.
\end{align}
The equation above solves for partial batches that maximizes total prefill budgets while attaining decode SLOs in a time interval $t$.
A partial batch specifies its total token budget, allocates them to decode requests and leaves the rest as prefill budget.
The partial batch's total token budget and decode token allocation can be used by the performance model in \S\ref{performance model} to estimate the per-batch execution time.
Using the performance model, a solver identifies the optimal list of partial batches that maximize the remaining prefill budgets after satisfying the decode SLOs of accepted requests.
We later present two solvers: one for auto-regressive decoding based serving (Algorithm~\ref{alg:batch formation}), and another for speculative-decoding based serving (\S\ref{sec:slo-ada-sd}).

\paragraph{The Optimal Solution.} The optimal value $v^*$ is identified by $(i^*, n^*), v^* = \operatorname{arg}\max_{i,n}DP[i,M,0,n]$, where $M$ denotes total memory units. Here, $R_{i^*}$ is the last accepted request, and $n^*$ is the total number of accepted requests. 
We reconstruct the optimal schedule by tracking $j^*[i^*,M,0,n^*]$ from Eqn.~\ref{eqn:dp} for admission decisions, and $B^*[i^*,M,0,n^*]$ from Eqn.~\ref{eqn:pb} for the scheduled partial batches. 
Since tokens in partial batches are only allocated to decode requests, we allocate the prefill budget in every batch by prioritizing requests with earlier prefill deadlines. 
It is guaranteed to have sufficient prefill budgets for allocation, because the DP algorithm requires the prefill budget to be non-negative by every prefill deadline.

\paragraph{Continuous Optimization.} The scheduler, previously discussed for a new-request-only scenario, can be extended to accommodate running requests through forced admission.
Forcing admission for running requests maintains \SysName's invariant of guaranteeing SLOs for all admitted requests. 
Specifically, during enumeration in Eqn.~\ref{eqn:dp}, the last admitted request before $R_i$ is constrained to either the latest running request with a prior prefill deadline or any request within their respective deadlines.
Continuous optimization, as shown in \S\ref{sch-overhead}, minimizes the overhead imposed by running requests.

\paragraph{Multi-Decode SLOs.} For a request with multiple decode SLOs (e.g., a reasoning request with a tight SLO for thinking and a loose SLO for response), its tightest SLO is taken to upper-bound the resource demand. When requests have different decode SLOs, the DP scheduler tracks the accepted requests per TPOT SLO, updating its state to include SLO-specific request counts: $(i, m, pb, (n_1, n_2, ..., n_L))$. This information enables the solver to generate optimal batch schedules across varying decode SLOs.

\paragraph{Time Complexity} 
Given $N$ total requests, $N_{new}$ new requests, $L$ decode tiers, and $M$ memory units, since every transition enumerates at most $N_{new}$ previous states, the total time complexity is bounded by $O(|States|\times N_{new}) = O(N\cdot N_{new}^{L+1}\cdot M\cdot \Sigma_i p_i)$. When the optimization goal is request throughput under SLOs, the time complexity is reduced to $O(N\cdot N_{new}^{L+1}\cdot M)$ (see Appendix~\ref{apx:time-complexity}). 
 In practice, with typical workloads of zero to ten new requests and dozens to hundreds of running requests, our scheduler introduces negligible overhead (see more in \S\ref{sch-overhead}).

\subsubsection{Batch Formation with Dynamic Size Tuning}\label{sec:bs-tuning}

We now describe our approach to form batches in Eqn.~\ref{eqn:pb} with auto-regressive decoding (top row in Figure~\ref{sec:slo-ada-sd}) that maximizes the prefill token budgets subjecting to the decode SLO constraints.
Since token throughput increases monotonically with batch sizes (Fig.~\ref{fig:batching}), we form batches by \textit{selecting the largest possible batch size that satisfies the decoding SLOs of all running requests.} 
Algorithm~\ref{alg:batch formation} details this process. First, the tightest TPOT SLO among running requests is determined (line~\ref{line:tighestTPOT}), which sets the latency for all scheduled batches. Subsequently, tokens are allocated to decoding requests in an early deadline first priority queue (lines~\ref{line:allocBeg}-\ref{line:allocEnd}), ensuring their SLOs are met. Compared to Sarathi-Serve, which globally caps batch sizes based on the tightest TPOT SLO, this algorithm improves token throughput by dynamically adapting batch sizes to the current set of running requests.

\begin{algorithm}[htbp]
\caption{Batch Formation}
\label{alg:batch formation}
\KwIn{$t$, $Reqs_{Decoding}$, $\mathcal{M}: \text{Perf. Model}$}
\KwOut{$Batches$}
\BlankLine
$t_0 \leftarrow \min_{req \in Reqs_{Decoding}} req.TPOT$ \label{line:tighestTPOT}\;
\For{$req \in Reqs_{Decoding}$}{$req.schDDL \leftarrow 0$ \;}
$Q \leftarrow \text{PrioQueue}(Reqs_{Decoding}, \text{schDDL})$ \label{line:allocBeg}\;
$batches \leftarrow []$ \;
\For{$i \leftarrow 0 ... \lfloor t/t_0 \rfloor - 1$}{
    $b \leftarrow \text{Batch}(prefillBgt=\mathcal{M}.time2bs(t_0))$ \;
    \While{$Q.\text{front()}.schDDL \geq i \cdot t_0 \wedge b.prefillBgt > 0$}{
        $req \leftarrow Q.\text{pop}()$ \;
        $b.\text{scheduleDecode}(req)$ \;
        $b.prefillBgt \leftarrow b.prefillBgt - 1$ \;
        $req.schDDL \leftarrow req.schDDL + req.TPOT$ \;
        $Q.\text{push}(req)$ \;
    }
    $batches.\text{append}(batch)$\label{line:allocEnd} \;
}
\textbf{return} $batches$ \;
\end{algorithm}

\subsubsection{SLO adaptive Speculative Decoding}\label{sec:slo-ada-sd}

\begin{figure}
    \centering
    \includegraphics[width=\linewidth]{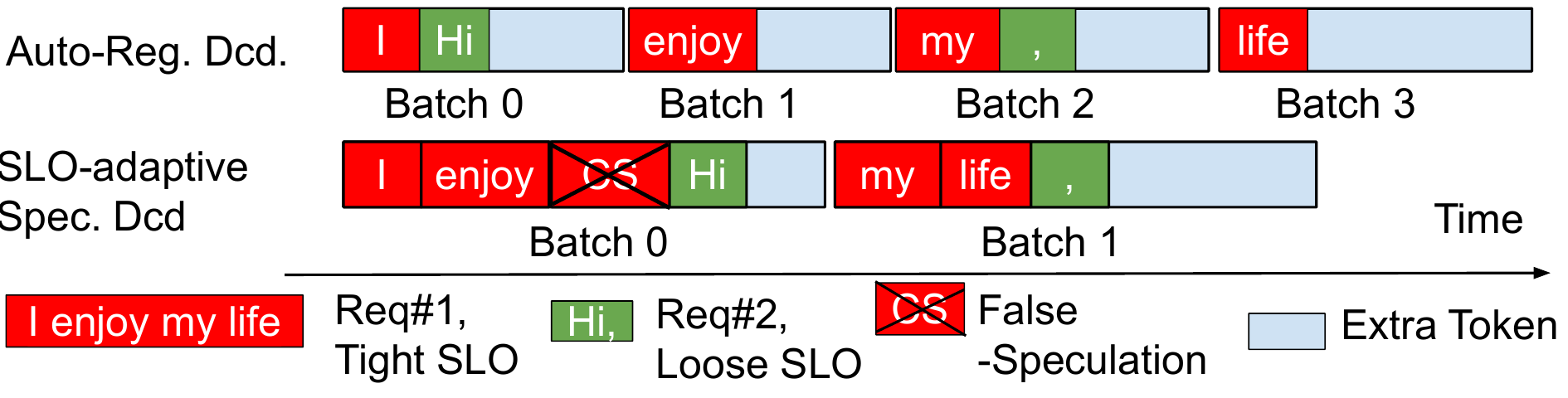}
    \caption{SLO-adaptive Speculative Decoding}
    \label{fig:slo-adaptive speculative decoding}
\end{figure}
We extend the batch formation technique to incorporate speculative decoding for enhanced efficiency.
With speculative decoding, token throughput is increased by relieving the per-batch latency constraint through processing more tokens per batch.
This is because, as illustrated in Fig.~\ref{fig:slo-adaptive speculative decoding}, auto-regressive decoding constrains batch token capacity based on the tightest decode SLO.
In contrast, speculative decoding permits the processing of multiple decode tokens per request within a batch, thereby relaxing per-batch latency constraints (e.g., $2\times TPOT$ when generating two tokens per batch) and increasing token throughput. Furthermore, with multiple decode SLOs, token allocation can be dynamically adjusted per request, enabling \SysName to maximize the prefill token budget within a defined time frame. A contemporary work~\cite{li2025adaserve} proposes a similar idea to customize token allocation in speculative decoding for multi-decode SLOs. However, they are focusing on a decode-only setup, while \SysName targets multi-stage scenarios and utilizes speculative decoding to enable higher token throughput.

Specifically, suppose there are $n_1, ..., n_L$ requests in the decoding stage running with TPOTs $TPOT_1, ..., TPOT_L$, we decide the batch in Eqn.~\ref{eqn:pb} with the maximum prefill token throughput by solving the following problem (see details in Appendix.~\ref{apx:ada-spec}):
\begin{align*}
\max_{\substack{sl_{1:L}}} prefillTpt &:= \frac{\text{PrefillBgtPerBatch}}{\text{Batch Time}}  \\
\text{PrefillBgtPerBatch} &= Time2BS(T(sl_{1:L}), sl_{1:L}) - \sum_i n_i sl_i \\
\text{Batch Time} &= T(sl_{1:L}) = \min_{l \in 1, 2, ..., L}(TPOT_l \cdot Acc(sl_l)) 
\end{align*}
Here, $sl_{1:L}$ are per-decode SLO speculation lengths, and $Acc(sl_l)$ is the expected number of tokens generated for requests with $TPOT_l$. Then, the optimal speculation lengths and prefill budget are used to construct batches used in the DP algorithm.

 
To account for the uncertainty in speculative decoding's prediction, we dynamically adjust the request's decode SLOs. For example, we strengthen its SLO when a request falls behind its SLO in the decoding stage.

\section{\SysName}\label{S:application}
Building on top of the soft admission control mechanism, we design \SysName with two fallback mechanisms that enable \textit{burst resilient scheduling} and \textit{SLO-driven request routing}.

\subsection{Burst Resilient Scheduling}\label{sec:burst}

When request bursts occur, resulting in instantaneous loads exceeding server capacity, \SysName introduces \textit{burst resilient scheduling}. This mechanism delays requests with unat\-tainable SLOs to ensure SLO attainment for the remaining requests.

In addition to standard services with defined SLOs, \SysName incorporates a cost-effective best-effort service tier. This tier utilizes any leftover resources after fulfilling SLO-guaranteed requests, similar to OpenAI's batch API~\cite{openai_batch_overview}, which offers lower costs with a longer response time. The scheduler seamlessly integrates this best-effort tier by batching requests to consume surplus token budgets and executing them when sufficient memory is available. To prioritize SLO-guaranteed requests, the scheduler can preempt best-effort requests upon new arrivals. Preemption overhead is minimized by discarding only the cached states (KV Cache), while retaining the generated tokens. This allows preempted requests to resume with a single prefill step, recomputing the cache for the original prompt and previously generated tokens, rather than repeating the entire decoding process.

\SysName then employs the best-effort service tier to handle bursts and requests with unattainable SLOs. After the admission decision by the scheduler, requests with unattainable SLOs are offloaded to the best-effort tier. While this tier lacks SLO guarantees, it is shown later that these requests are efficiently handled during low or zero-load periods following the burst (Fig.~\ref{fig:load-across-time}), maximizing system utilization.

\begin{figure}[t]
    \centering
     \includegraphics[width=0.75\linewidth]{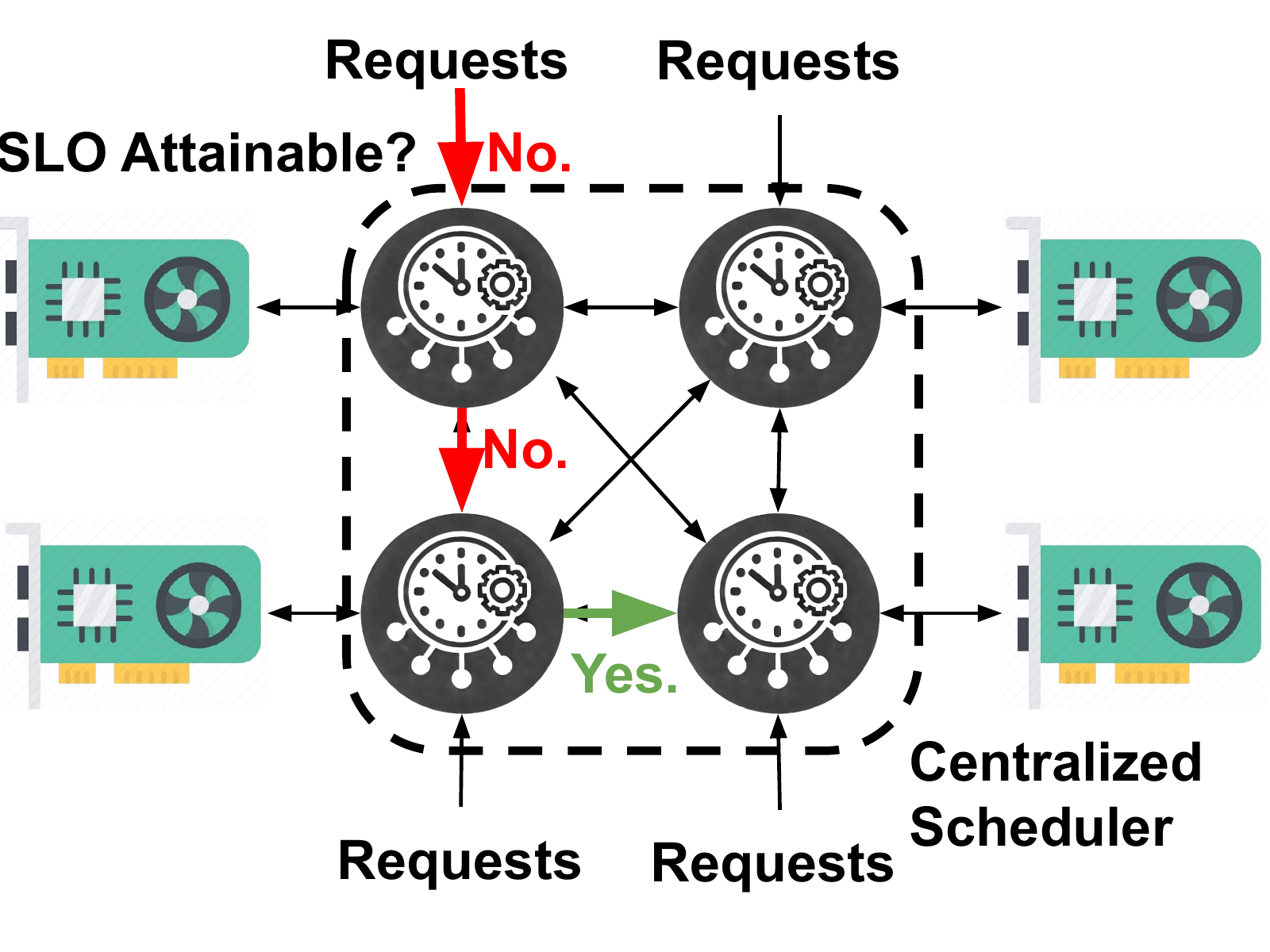}
    \caption{\SysName's Multi-Replica Serving.}
    \label{fig:multi-replica-serving}
\end{figure}

\subsection{Multi-replica Serving with SLO-driven Request Routing}\label{sec:multi-replica}

\SysName's scheduler and performance modeling provide the foundation for a scalable multi-replica LLM serving system with high per-GPU utilization.

Fig.~\ref{fig:multi-replica-serving} shows the architecture of \SysName's multi-replica serving framework. Each replica is associated with a scheduler operating within a centralized controller, which virtualizes execution using the performance model. Upon arrival of a new request at a replica (dispatched by a load balancer), the scheduler determines SLO attainability. If the current replica cannot meet the SLO, the request is routed to the next replica sequentially. When the routing counts reach a pre-configured limit, a backup policy is invoked, deciding whether to decline the request or offload it to lower-tier resources. While previous works on LLM load balancing always rely on device profiling with indirect metrics (e.g., device loads and memory consumption)~\cite{sun2024llumnix,stojkovic2024dynamollm,hu2024inference}, \SysName directly balances the load based on SLO attainment, looking forward to full GPU utilization and rigorous SLO attainment. Note that sequential routing is one of many ways~\cite{gawlick1995admission} to leverage \SysName's admission control mechanism in multi-replica routing, and we leave the exploration of this to future work.

\section{Implementation}
We build \SysName as a distributed LLM serving system that supports multi-SLOs, multi-replica serving, while being resilient to request bursts. 
Currently, we support vLLM as our backend but it is easy to extend to other backends as long as they comply with our batched execution model.
We utilize PagedAttention~\cite{vllm} for memory management and Ray~\cite{moritz2018ray} for the orchestration between the scheduler and the backend. The implementation of the scheduling orchestrations with the executor interface, serving frontend, takes 10K lines of Python Code, whereas the resource planning algorithm is highly crafted in C++ with 1.5K LoC.



\section{Evaluation}\label{sec:eval}

We evaluate \SysName on a total of six scenarios (Tab.~\ref{tab:scenario}) and five datasets (Tab.~\ref{tab:datasets}). We consider vLLM, Sarathi-Serve, DistServe as our baselines as they represent SOTA for LLM serving. Our evaluation seeks to answer these questions:

\begin{asparaitem}
    \item What is the maximum capacity (i.e., request load) a single GPU can serve under specific SLO constraints with different serving systems, and how well does \SysName stay resilient with bursty arrivals (\S\ref{sec:end2end})?
    \item In multi-replica serving, how does \SysName scales with the number of replicas (\S\ref{sec:scaling})?
    \item How do \SysName's individual optimizations in isolation contribute to its performance (\S\ref{sec:ablation})?
\end{asparaitem}

\begin{table}[t]
\caption{Experiment Scenarios.}
\label{tab:scenario}
\resizebox{\linewidth}{!}{
\begin{tabular}{@{}l|c|c|c@{}}
\toprule
\textbf{Scenarios} & \textbf{Arrival Pattern} & \textbf{Prompt Dataset} & \textbf{Model} \\ 
\midrule
ChatBot & Azure-Chatting~\cite{patel2024splitwise} & ShareGPT~\cite{sharegpt} & \multirow{2}{*}{Main: OPT-7B/13B/30B~\cite{zhang2022opt}}  \\ 
Coder & Azure-Coding~\cite{patel2024splitwise} & HumanEval~\cite{chen2021codex} &  \\ 
Summarizer & Azure-Chatting & Arxiv Summary~\cite{cohan2018discourse} & Spec: OPT-125m \\ 
\midrule
Mixed & \multicolumn{3}{c}{ChatBot + Coder + Summarizer} \\ 
\midrule
ToolLLM & Azure-Coding & ToolBench~\cite{guo2024stabletoolbench} & ToolLlama-7B~\cite{qin2023toolllm} \\ 
\midrule
Reasoning & Azure-Chatting & s1K~\cite{muennighoff2025s1} & Deepseek-R1-Qwen-1.5B~\cite{guo2025deepseek} \\
\bottomrule
\end{tabular}
}
\end{table}

\paragraph{Setup.} We evaluate \SysName across different models: OPT series~\cite{zhang2022opt} for standard scenarios (ChatBot, Coder, Summarizer), ToolLlama~\cite{qin2023toolllm} for ToolLLM, and a distilled Deepseek-R1-Qwen model~\cite{guo2025deepseek} for the reasoning scenario---these models are either used in evaluation by previous work~\cite{leviathan2023fast,zhong2024distserve} or well-suited for the application.
For the end-to-end capacity evaluation, we use the Google a2-highgpu-4g VM with 4 NVIDIA 40GB A100 GPUs, connected with pairwise NVLINK. For the OPT-13B (OPT-30B) model, we run with 2-way (4-way) tensor parallelism. For the OPT series, we use OPT-125m as the speculative decoding model and replicate it on every GPU.
For the OPT-7B model, we run \SysName with up to 4 replicas.
For other models, we run \SysName with a single replica.
We run \SysName with a best-effort tier across all scenarios.
For memory management, we assume prior knowledge on decode lengths available to all baselines, which can be well-approximated from previous works~\cite{sun2024llumnix,qin2024mooncake,hu2024inference,prabhu2024vattention}.
For the scalability evaluation, we use an additional Google a3-highgpu-8g with 8 NVIDIA 80GB H100 GPU to perform multi-replica serving for the OPT-13B model.


\begin{figure}
    \centering
    \begin{subfigure}[b]{0.48\columnwidth} 
        \includegraphics[width=\linewidth]{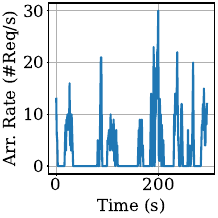}
        \caption{Coding Trace.}
        \label{trace:code}
    \end{subfigure}
    \begin{subfigure}[b]{0.48\columnwidth} 
        \includegraphics[width=\linewidth]{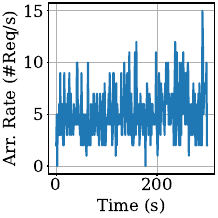}
        \caption{Chatting Trace.}
        \label{trace:chat}
    \end{subfigure}
    \caption{Traces from Azure Dataset~\cite{patel2024splitwise} used in evaluation.}
    \label{fig:trace}
\end{figure}


\paragraph{Workloads} We used Azure LLM inference traces~\cite{patel2024splitwise} for realistic request arrival patterns: Coding shows bursty arrivals, while Chatting is more stable (Fig.~\ref{fig:trace}). Request lengths are based on Azure traces, supplemented by Arxiv Summary~\cite{cohan2018discourse}, ToolBench~\cite{guo2024stabletoolbench}, and s1K~\cite{muennighoff2025s1} for specific scenarios (Tab.~\ref{tab:datasets}). ChatBot and Reasoning are decode-heavy with long outputs, Summarizer is prefill-heavy, and Coder and ToolLLM exhibit the largest input/output length variations.

\paragraph{SLOs.} We enforce the SLO based on the specification in Tab.~\ref{tab:stages}. Particularly, we focus on the max TTFT slowdown compared to zero-load setup for the prefill stages and max TPOT for the decode stages. Because speculative decoding outputs more than one token at a time, we measure the TPOT every 10 tokens.

\paragraph{Baseline} We consider vLLM~\cite{vllm} (commit \texttt{c38eba30}), Sarathi-Serve~\cite{sarathi} (enabled with vLLM), and DistServe~\cite{zhong2024distserve} (commit \texttt{0a2d0c99}) as our baseline as they represent the SOTA serving frameworks. For vLLM, we additionally evaluate its speculative decoding version (vLLM (Spec)). For Sarathi-Serve, we configure the batch size to the maximum size without violating the tightest decode SLO. For DistServe, we experimented with different prefill-decode device allocations (1:1, 2:1, 1:2) and report the best result.

\paragraph{Metric} We evaluate serving capacity (maximum GPU request load with less than 10\% SLO violations~\cite{zhong2024distserve,sarathi}). For multi-GPU systems, we normalize the request rate by dividing the total request rate by the number of GPUs.

\begin{table}[h]
\centering
\caption{SLOs for different model configurations.}
\label{tab:SLOs}
\begin{tabular}{c|cc}
\toprule
           & Max TTFT Slowdown & Max TPOT  \\ 
\midrule
Tight SLO & 3x                 & 50ms  \\
Loose SLO & 5x                & 100ms  \\
\bottomrule
\end{tabular}
\end{table}

\begin{table}[h]
    \centering
    \caption{Datasets used for evaluation. For the reasoning task, we list the number of tokens for thinking/response. For the ToolLLM, there is 2.7 +- 1.1 prefill-decode pairs in a request.}
    \label{tab:datasets}
    \resizebox{\columnwidth}{!}{
\begin{tabular}{l|ccc|ccc}
\toprule
 & \multicolumn{3}{c}{\textbf{Prompt Tokens}} & \multicolumn{3}{c}{\textbf{Output Tokens}} \\
 & Mean & P99 & Std. & Mean & P99 & Std. \\
\midrule
ChatBot & 763 & 1591 & 424 & 266 & 619 & 160 \\
Coder & 847 & 2010 & 617 & 26 & 232 & 47 \\
Reasoning & 127 & 421 & 83 & 4693/803 & 7297/1650 & 1442/280 \\
Summarizer & 1333 & 1946 & 444 & 202 & 1508 & 234 \\
ToolLLM & 690 & 2131 & 356 & 116 & 363 & 66 \\
\bottomrule
\end{tabular}
    }
\end{table}




\begin{figure*}
    \includegraphics{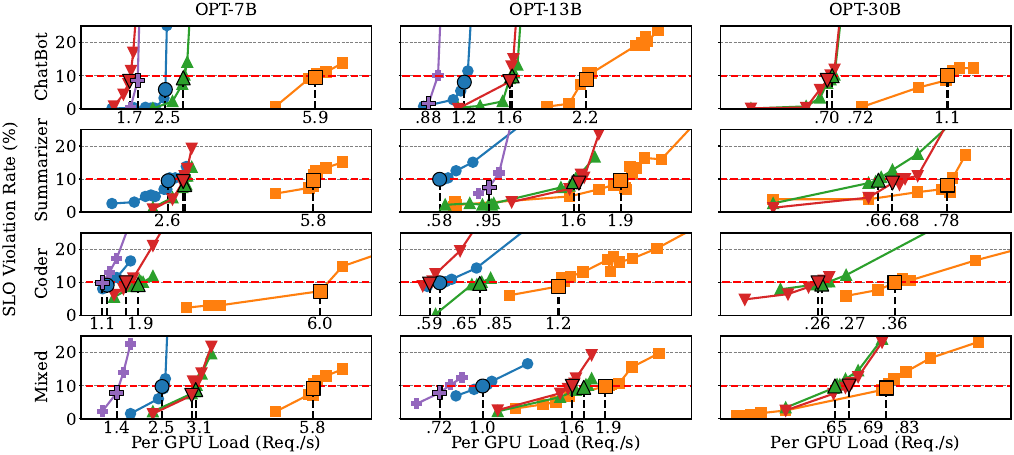}
    \hfill
    \includegraphics{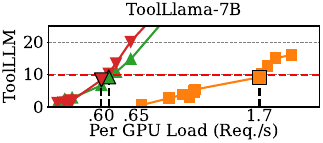}
    \includegraphics{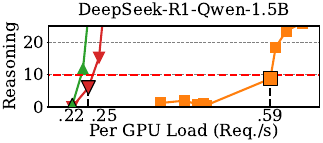}
    \includegraphics{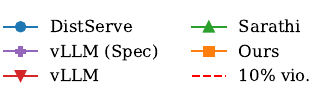}
    \caption{End-to-End evaluation of \SysName.}
    \label{fig:e2e}
\end{figure*}

\begin{figure}
    \centering
    \begin{subfigure}[t]{1.65in}
    \includegraphics[width=\textwidth]{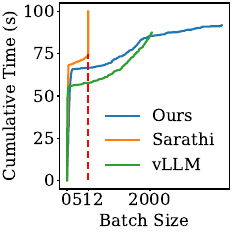}
    \caption{Cumulative Execution Time along Batch Sizes for the OPT-7B with Summarization Scenario at 3.0 request/s.}
    \label{fig:batch_size}
    \end{subfigure}
    \hfill
    \begin{subfigure}[t]{1.65in}
    \includegraphics[width=\textwidth]{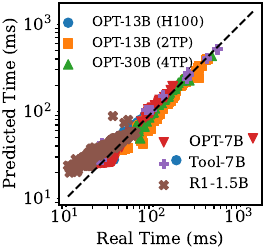}
    \caption{Predicted Time vs. Real Time for the performance model across LLMs, \#Tensor Parallel (TP), and hardwares (A100 if unmarked).}
    \label{fig:perf-model}
    \end{subfigure}
    \caption{Analysis.}
\end{figure}

\subsection{End-to-end Capacity Evaluation}\label{sec:end2end}

The end-to-end evaluation (Fig.~\ref{fig:e2e}) demonstrates that \SysName consistently outperforms baseline systems. Specifically, it achieves a 1.76x capacity improvement over vLLM, 1.94x over Sarathi, and 2.6x over DistServe. The performance gain is attributed to \SysName's optimizations, including SLO-optimized, burst-resilient scheduling, as well as request routing. Even when enforcing a stringent 2\% SLO violation constraint, which mitigates the impact of delaying requests with unattainable SLOs, \SysName maintains a capacity advantage in the ChatBot scenario: 1.91x over vLLM, 2.19x over Sarathi, and 2.4x over DistServe. These results underscore \SysName's robust support for multi-stage SLOs.

With stable request arrival rates in both ChatBot and Summarizer scenarios (Figure~\ref{fig:e2e}), maximizing token throughput under multi-stage SLOs is the key to high serving capacity.
Among baselines, vLLM prioritizes prefill, resulting in widespread decode SLO violations. \SysName and Sarathi multiplex stages, but Sarathi's fixed batch size limits throughput, particularly in the Summarizer scenario with longer input text and tight prefill SLOs. \SysName addresses this with SLO-adaptive speculative decoding and dynamic batch size tuning, enabling larger batches. Specifically, shown in Figure~\ref{fig:batch_size}, \SysName utilizes batches exceeding 512 tokens for up to 25\% of execution time, while Sarathi is capped at 512. Consequently, \SysName achieves 1.41-1.70x and 1.14-1.17x serving capacity than baselines across OPT-7B, 13B, and 30B models for ChatBot and Summarizer, respectively.

\begin{figure}[h!]
    \centering
    \centering
      \includegraphics[width=0.95\linewidth]{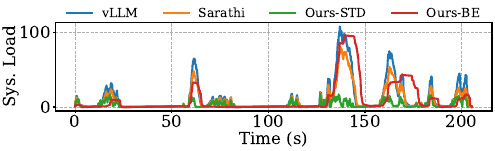}
    \caption{System Load (\# Req. in System) across time for Coder under high-load (4.5 Req/s) scenario. Ours separates into standard services (STD) and the best effort service (BE).}
    \label{fig:load-across-time}
\end{figure}

\begin{figure}
    \begin{subfigure}[t]{\linewidth}
    \includegraphics[width=\linewidth]{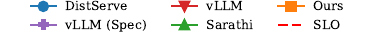}
    \end{subfigure}
    \begin{subfigure}[t]{0.45\linewidth}
        \centering
        \includegraphics[width=\textwidth]{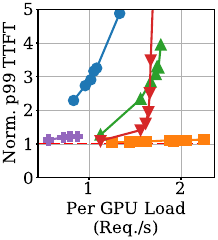}
    \end{subfigure}
    \hfill
    \begin{subfigure}[t]{0.45\linewidth}
    \includegraphics[width=\textwidth]{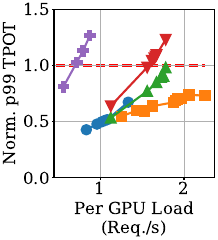}
    \end{subfigure}
    \caption{TTFT/TPOT Comparison for Mixed Scenario serving 13B model. Ours shows the requests belonging to the standard service tier.}
    \label{fig:mixed}
\end{figure}

\begin{figure}[!t]
    \centering
    \includegraphics[width = \linewidth]{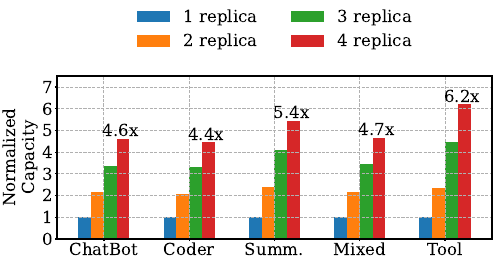}
    \caption{Capacity Scaling in \SysName's multi-replica Serving on 4 A100 GPUs with OPT-7B model.}
    \label{fig:scaling}
\end{figure}

Coder and ToolLLM scenarios feature bursty request arrivals (Figure~\ref{trace:code}) and short outputs, highlighting \SysName's burst-resilient scheduling. \SysName effectively manages load spikes (Figure~\ref{fig:load-across-time}) by deferring requests with unattainable SLOs (red line) during surges and processing them in low-load periods, maintaining SLO attainment for others (green line). Baselines, lacking this capability, suffer from cumulative delays during bursts, leading to violations of all request SLOs. Consequently, \SysName achieves up to 2.1x and 1.90x serving capacity than baselines in Coder and ToolLLM scenarios (Figure~\ref{fig:e2e}, third row and leftmost figure on the last row), respectively. This demonstrates robust burst handling, especially notable in ToolLLM where \SysName operates without the benefit of a speculative decoding model.


The Mixed scenario simulates diverse application workloads, testing system adaptability to varied patterns and SLOs. Shown in Figure~\ref{fig:mixed}, at 1.5 requests/second, vLLM and Sarathi exhibit significant p99 TTFT degradation, exceeding SLOs. Conversely, \SysName admission control mechanism maintains p99 TTFT and TPOT near specified SLOs for the standard services. 

Lastly, the reasoning scenario has long generation lengths with customized SLOs for the thinking stage and the decode stage.
The longer per-request lifespans indicate more concurrent requests in the system, magnifying the stalls caused by greedy schedulers.
In contrast, \SysName's SLO-optimized scheduling algorithm ensures multi-stage SLO attainment, as well as improving throughput by tuning the batch sizes to fit the per-request SLOs. As a result, even without a speculative model, \SysName supports 2.4x serving capacity compared to baselines (Figure~\ref{fig:e2e}, second on the last row). 

\subsection{Scaling Evaluation}\label{sec:scaling}

Now, we evaluate \SysName's multi-replica on the cluster with 4 A100 GPUs, serving the OPT-7B model with up to 4 replicas.
Figure~\ref{fig:scaling} shows the serving capacity across five scenarios.
Following the mechanism in Figure~\ref{fig:multi-replica-serving}, a request is first routed to a replica in by a one-shot round-robin dispatcher.
Afterwards, depending on the admission decision of \SysName's scheduler for that replica, the request either gets served or routed to the next replica.

Across five cases, \SysName exhibits linear or higher scaling. For example, in the ChatBot scenario, \SysName sustains 2.18x, 3.36x, and 4.61x higher capacity with 2,3,4 replicas when serving an OPT-7B model. 

\SysName's super-linear scalability, while seems counter-intuitive, is the result of its scheduling mechanism (Algorithm~\ref{alg:scheduler}). Although \SysName's scheduler makes locally optimal admission decisions upon each invocation, it is inherently greedy to currently present requests. When scaling to a multi-replica configuration, while each individual replica's scheduler maintains a constant request rate, the aggregate of all schedulers across replicas observes a linearly increasing number of requests. This expanded global view, facilitated by request routing, enables the combined scheduler network to make more informed, globally optimized admission decisions. Consequently, in scenarios like ToolLLM and Coder, where requests exhibit significant variance in input/output lengths, this multi-replica serving strategy mitigates the limitations of local greedy decisions, resulting in a 5.44x and 6.2x increase in served load on four replicas, respectively.

\begin{figure}[t]
    \centering
    \includegraphics{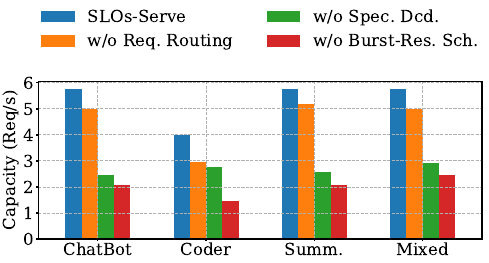}
    \caption{Ablation Study. We measure the serving capacity by gradually removing \SysName's optimizations.}
    \label{fig:ablation}
\end{figure}

\begin{figure}[t]
    \centering
    \includegraphics{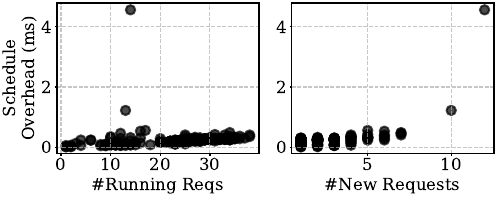}
    \caption{Schedule Overhead.}
    \label{fig:runtime}
\end{figure}



\subsection{Ablation Study}\label{sec:ablation}

We conducted an ablation study (Figure~\ref{fig:ablation}) to quantify the individual contributions of \SysName's optimizations. Starting from full \SysName, we remove multi-replica serving with request routing, SLO-adaptive speculative decoding, and the burst-resilient scheduling across four scenarios. For the baseline case, we implemented a prefill-oriented scheduling approach in \SysName for apple-to-apple comparison.

We found that request routing, SLO-adaptive speculative decoding, and burst-resilient serving independently improve serving capacity by 1.19x, 1.66x, and 1.34x. 
SLO-adaptive decoding demonstrated its largest gains in decode-heavy workloads (2.01x  for ChatBot and 2.02x for Summarizer) by enabling larger token batch sizes.
Conversely, request routing (1.92x) and burst-resilient serving (1.66x) were most effective in the bursty, variable-length Coder scenario, where they balanced instance load and deferred request with unattainable SLOs to secure the SLOs of the rest, respectively.

\subsection{Scheduling Overhead}\label{sch-overhead}

Last but not least, we assess the scheduling overhead of \SysName's scheduling algorithm. As depicted in Figure~\ref{fig:runtime}, the scheduling overhead for each call to \SysName's resource planning algorithm, profiled from real scheduling scenarios, consistently remains under 10 milliseconds, with the majority of overheads concentrated below 2 milliseconds. Considering that the scheduler typically plans 1-5 future batches, and each batch requires at least 25 milliseconds to complete, the overhead introduced by the scheduling algorithm is minimal. This ensures that the low-overhead request routing is in multi-replica serving.

\subsection{Fidelity of the performance model}
To validate our performance model, we conducted evaluations across diverse configurations (models, tensor parallel setups, speculative decoding, hardware architecture). Figure~\ref{fig:perf-model} shows that the model consistently achieved high fidelity, with R-squared scores ranging from 0.82 to 0.93.
\section{Related Work}

\paragraph{LLM Serving System.}

There are a variety of recent works on serving Large Language Models (LLM). These works have diverse research focus, including optimizing attentions~\cite{vllm,vattention,loongserve}, prefill/decode disaggregation~\cite{zhong2024distserve,patel2024splitwise}, KV cache~\cite{qin2024mooncake,zheng2024sglang,hu2024memserve}, heterogeneous serving~\cite{helix}, on-device serving~\cite{powerinfer,neupims}, serverless serving~\cite{medusa}, and serving emerging LLM model variations~\cite{moe_lightning,apparate}.
A few previous works focus on similar goals (optimizing for SLOs) as \SysName, including Llumnix~\cite{sun2024llumnix}, DynamoLLM~\cite{stojkovic2024dynamollm}, ExeGPT~\cite{exegpt} and VTC~\cite{sheng2024fairness}. Comparing to these works, \SysName uses admission control paired with a dynamic programming alaogirithm to guarantee attainment of multi-stage SLOs.

\paragraph{Admission Control.}
\textit{Admission control} is often needed when the server is under high request load.
By rejecting some incoming requests, admission control ensures SLOs of running requests are satisfied.
In the literature, admission control is used in cloud services~\cite{sajal2023kerveros,gawlick1995admission}, elastic training~\cite{gu2023elasticflow}. Along serving frameworks like vLLM~\cite{vllm} may decline requests when the memory is not sufficient, \SysName admission control directly focuses on SLO-attainment.

\section{Conclusion}

In this work, we built \SysName, a distributed LLM serving system that supports multi-SLOs and multi-replica serving, while being resilient to burst arrivals. 
Particularly, we solve the key challenge of serving under application- and stage-dependent SLOs in a resource-sharing environment by customizing token allocations in batches.
\SysName then designs a dynamic programming-based algorithm that effectively detects requests with unattainable SLOs and continuously optimizes the token compositions by exploring the full design space of chunked prefill and (optional) speculative decoding. 
Built on top of the scheduler, \SysName enabled burst handling and low overhead routing in multi-replica serving.
In evaluation, comprehensive studies across six applications showed on average  2.2x improvement in serving capacity. In multi-replica serving, \SysName embodied super-linear scaling with bursty arrivals, underscoring the effectiveness of dynamic request routing.

\section{Acknowledgement}
This work is partly supported by grants from the National Science Foundation (NSF CNS-2211882), and by the member companies of the Wasm Research Center and PDL consortium. We thank Juncheng Gu for precious comments on the paper.

\bibliographystyle{ACM-Reference-Format}

\begin{thebibliography}{41}


\ifx \showCODEN    \undefined \def \showCODEN     #1{\unskip}     \fi
\ifx \showISBNx    \undefined \def \showISBNx     #1{\unskip}     \fi
\ifx \showISBNxiii \undefined \def \showISBNxiii  #1{\unskip}     \fi
\ifx \showISSN     \undefined \def \showISSN      #1{\unskip}     \fi
\ifx \showLCCN     \undefined \def \showLCCN      #1{\unskip}     \fi
\ifx \shownote     \undefined \def \shownote      #1{#1}          \fi
\ifx \showarticletitle \undefined \def \showarticletitle #1{#1}   \fi
\ifx \showURL      \undefined \def \showURL       {\relax}        \fi
\providecommand\bibfield[2]{#2}
\providecommand\bibinfo[2]{#2}
\providecommand\natexlab[1]{#1}
\providecommand\showeprint[2][]{arXiv:#2}

\bibitem[sha({[n.\,d.]})]%
        {sharegpt}
 \bibinfo{year}{[n.\,d.]}\natexlab{}.
\newblock \bibinfo{title}{ShareGPT}.
\newblock \bibinfo{howpublished}{\url{https://sharegpt.com/}}.
\newblock
\newblock
\shownote{Accessed: 2024-12-07}.


\bibitem[Agrawal et~al\mbox{.}(2024)]%
        {sarathi}
\bibfield{author}{\bibinfo{person}{Amey Agrawal}, \bibinfo{person}{Nitin
  Kedia}, \bibinfo{person}{Ashish Panwar}, \bibinfo{person}{Jayashree Mohan},
  \bibinfo{person}{Nipun Kwatra}, \bibinfo{person}{Bhargav~S Gulavani},
  \bibinfo{person}{Alexey Tumanov}, {and} \bibinfo{person}{Ramachandran
  Ramjee}.} \bibinfo{year}{2024}\natexlab{}.
\newblock \showarticletitle{Taming throughput-latency tradeoff in llm inference
  with sarathi-serve}.
\newblock \bibinfo{journal}{\emph{arXiv preprint arXiv:2403.02310}}
  (\bibinfo{year}{2024}).
\newblock


\bibitem[Cao et~al\mbox{.}(2025)]%
        {moe_lightning}
\bibfield{author}{\bibinfo{person}{Shiyi Cao}, \bibinfo{person}{Shu Liu},
  \bibinfo{person}{Tyler Griggs}, \bibinfo{person}{Peter Schafhalter},
  \bibinfo{person}{Xiaoxuan Liu}, \bibinfo{person}{Ying Sheng},
  \bibinfo{person}{Joseph~E. Gonzalez}, \bibinfo{person}{Matei Zaharia}, {and}
  \bibinfo{person}{Ion Stoica}.} \bibinfo{year}{2025}\natexlab{}.
\newblock \showarticletitle{MoE-Lightning: High-Throughput MoE Inference on
  Memory-constrained GPUs}. In \bibinfo{booktitle}{\emph{Proceedings of the
  30th ACM International Conference on Architectural Support for Programming
  Languages and Operating Systems, Volume 1}} (Rotterdam, Netherlands)
  \emph{(\bibinfo{series}{ASPLOS '25})}. \bibinfo{publisher}{Association for
  Computing Machinery}, \bibinfo{address}{New York, NY, USA},
  \bibinfo{pages}{715–730}.
\newblock
\showISBNx{9798400706981}
\href{https://doi.org/10.1145/3669940.3707267}{doi:\nolinkurl{10.1145/3669940.3707267}}


\bibitem[Chen et~al\mbox{.}(2021)]%
        {chen2021codex}
\bibfield{author}{\bibinfo{person}{Mark Chen}, \bibinfo{person}{Jerry Tworek},
  \bibinfo{person}{Heewoo Jun}, \bibinfo{person}{Qiming Yuan},
  \bibinfo{person}{Henrique~Ponde de Oliveira~Pinto}, \bibinfo{person}{Jared
  Kaplan}, \bibinfo{person}{Harri Edwards}, \bibinfo{person}{Yuri Burda},
  \bibinfo{person}{Nicholas Joseph}, \bibinfo{person}{Greg Brockman},
  \bibinfo{person}{Alex Ray}, \bibinfo{person}{Raul Puri},
  \bibinfo{person}{Gretchen Krueger}, \bibinfo{person}{Michael Petrov},
  \bibinfo{person}{Heidy Khlaaf}, \bibinfo{person}{Girish Sastry},
  \bibinfo{person}{Pamela Mishkin}, \bibinfo{person}{Brooke Chan},
  \bibinfo{person}{Scott Gray}, \bibinfo{person}{Nick Ryder},
  \bibinfo{person}{Mikhail Pavlov}, \bibinfo{person}{Alethea Power},
  \bibinfo{person}{Lukasz Kaiser}, \bibinfo{person}{Mohammad Bavarian},
  \bibinfo{person}{Clemens Winter}, \bibinfo{person}{Philippe Tillet},
  \bibinfo{person}{Felipe~Petroski Such}, \bibinfo{person}{Dave Cummings},
  \bibinfo{person}{Matthias Plappert}, \bibinfo{person}{Fotios Chantzis},
  \bibinfo{person}{Elizabeth Barnes}, \bibinfo{person}{Ariel Herbert-Voss},
  \bibinfo{person}{William~Hebgen Guss}, \bibinfo{person}{Alex Nichol},
  \bibinfo{person}{Alex Paino}, \bibinfo{person}{Nikolas Tezak},
  \bibinfo{person}{Jie Tang}, \bibinfo{person}{Igor Babuschkin},
  \bibinfo{person}{Suchir Balaji}, \bibinfo{person}{Shantanu Jain},
  \bibinfo{person}{William Saunders}, \bibinfo{person}{Christopher Hesse},
  \bibinfo{person}{Andrew~N. Carr}, \bibinfo{person}{Jan Leike},
  \bibinfo{person}{Josh Achiam}, \bibinfo{person}{Vedant Misra},
  \bibinfo{person}{Evan Morikawa}, \bibinfo{person}{Alec Radford},
  \bibinfo{person}{Matthew Knight}, \bibinfo{person}{Miles Brundage},
  \bibinfo{person}{Mira Murati}, \bibinfo{person}{Katie Mayer},
  \bibinfo{person}{Peter Welinder}, \bibinfo{person}{Bob McGrew},
  \bibinfo{person}{Dario Amodei}, \bibinfo{person}{Sam McCandlish},
  \bibinfo{person}{Ilya Sutskever}, {and} \bibinfo{person}{Wojciech Zaremba}.}
  \bibinfo{year}{2021}\natexlab{}.
\newblock \showarticletitle{Evaluating Large Language Models Trained on Code}.
\newblock  (\bibinfo{year}{2021}).
\newblock
\showeprint[arxiv]{2107.03374}~[cs.LG]


\bibitem[Cohan et~al\mbox{.}(2018)]%
        {cohan2018discourse}
\bibfield{author}{\bibinfo{person}{Arman Cohan}, \bibinfo{person}{Franck
  Dernoncourt}, \bibinfo{person}{Doo~Soon Kim}, \bibinfo{person}{Trung Bui},
  \bibinfo{person}{Seokhwan Kim}, \bibinfo{person}{Walter Chang}, {and}
  \bibinfo{person}{Nazli Goharian}.} \bibinfo{year}{2018}\natexlab{}.
\newblock \showarticletitle{A discourse-aware attention model for abstractive
  summarization of long documents}.
\newblock \bibinfo{journal}{\emph{arXiv preprint arXiv:1804.05685}}
  (\bibinfo{year}{2018}).
\newblock


\bibitem[Dai et~al\mbox{.}(2024)]%
        {apparate}
\bibfield{author}{\bibinfo{person}{Yinwei Dai}, \bibinfo{person}{Rui Pan},
  \bibinfo{person}{Anand Iyer}, \bibinfo{person}{Kai Li}, {and}
  \bibinfo{person}{Ravi Netravali}.} \bibinfo{year}{2024}\natexlab{}.
\newblock \showarticletitle{Apparate: Rethinking Early Exits to Tame
  Latency-Throughput Tensions in ML Serving}. In
  \bibinfo{booktitle}{\emph{Proceedings of the ACM SIGOPS 30th Symposium on
  Operating Systems Principles}} (Austin, TX, USA) \emph{(\bibinfo{series}{SOSP
  '24})}. \bibinfo{publisher}{Association for Computing Machinery},
  \bibinfo{address}{New York, NY, USA}, \bibinfo{pages}{607–623}.
\newblock
\showISBNx{9798400712517}
\href{https://doi.org/10.1145/3694715.3695963}{doi:\nolinkurl{10.1145/3694715.3695963}}


\bibitem[Gawlick(1995)]%
        {gawlick1995admission}
\bibfield{author}{\bibinfo{person}{Rainer Gawlick}.}
  \bibinfo{year}{1995}\natexlab{}.
\newblock \showarticletitle{Admission control and routing: Theory and
  practice}.
\newblock  (\bibinfo{year}{1995}).
\newblock


\bibitem[Gu et~al\mbox{.}(2023)]%
        {gu2023elasticflow}
\bibfield{author}{\bibinfo{person}{Diandian Gu}, \bibinfo{person}{Yihao Zhao},
  \bibinfo{person}{Yinmin Zhong}, \bibinfo{person}{Yifan Xiong},
  \bibinfo{person}{Zhenhua Han}, \bibinfo{person}{Peng Cheng},
  \bibinfo{person}{Fan Yang}, \bibinfo{person}{Gang Huang},
  \bibinfo{person}{Xin Jin}, {and} \bibinfo{person}{Xuanzhe Liu}.}
  \bibinfo{year}{2023}\natexlab{}.
\newblock \showarticletitle{Elasticflow: An elastic serverless training
  platform for distributed deep learning}. In
  \bibinfo{booktitle}{\emph{Proceedings of the 28th ACM International
  Conference on Architectural Support for Programming Languages and Operating
  Systems, Volume 2}}. \bibinfo{pages}{266--280}.
\newblock


\bibitem[Guo et~al\mbox{.}(2025)]%
        {guo2025deepseek}
\bibfield{author}{\bibinfo{person}{Daya Guo}, \bibinfo{person}{Dejian Yang},
  \bibinfo{person}{Haowei Zhang}, \bibinfo{person}{Junxiao Song},
  \bibinfo{person}{Ruoyu Zhang}, \bibinfo{person}{Runxin Xu},
  \bibinfo{person}{Qihao Zhu}, \bibinfo{person}{Shirong Ma},
  \bibinfo{person}{Peiyi Wang}, \bibinfo{person}{Xiao Bi}, {et~al\mbox{.}}}
  \bibinfo{year}{2025}\natexlab{}.
\newblock \showarticletitle{Deepseek-r1: Incentivizing reasoning capability in
  llms via reinforcement learning}.
\newblock \bibinfo{journal}{\emph{arXiv preprint arXiv:2501.12948}}
  (\bibinfo{year}{2025}).
\newblock


\bibitem[Guo et~al\mbox{.}(2024)]%
        {guo2024stabletoolbench}
\bibfield{author}{\bibinfo{person}{Zhicheng Guo}, \bibinfo{person}{Sijie
  Cheng}, \bibinfo{person}{Hao Wang}, \bibinfo{person}{Shihao Liang},
  \bibinfo{person}{Yujia Qin}, \bibinfo{person}{Peng Li},
  \bibinfo{person}{Zhiyuan Liu}, \bibinfo{person}{Maosong Sun}, {and}
  \bibinfo{person}{Yang Liu}.} \bibinfo{year}{2024}\natexlab{}.
\newblock \bibinfo{title}{StableToolBench: Towards Stable Large-Scale
  Benchmarking on Tool Learning of Large Language Models}.
\newblock
\showeprint[arxiv]{2403.07714}~[cs.CL]


\bibitem[Heo et~al\mbox{.}(2024)]%
        {neupims}
\bibfield{author}{\bibinfo{person}{Guseul Heo}, \bibinfo{person}{Sangyeop Lee},
  \bibinfo{person}{Jaehong Cho}, \bibinfo{person}{Hyunmin Choi},
  \bibinfo{person}{Sanghyeon Lee}, \bibinfo{person}{Hyungkyu Ham},
  \bibinfo{person}{Gwangsun Kim}, \bibinfo{person}{Divya Mahajan}, {and}
  \bibinfo{person}{Jongse Park}.} \bibinfo{year}{2024}\natexlab{}.
\newblock \showarticletitle{NeuPIMs: NPU-PIM Heterogeneous Acceleration for
  Batched LLM Inferencing}. In \bibinfo{booktitle}{\emph{Proceedings of the
  29th ACM International Conference on Architectural Support for Programming
  Languages and Operating Systems, Volume 3}} (La Jolla, CA, USA)
  \emph{(\bibinfo{series}{ASPLOS '24})}. \bibinfo{publisher}{Association for
  Computing Machinery}, \bibinfo{address}{New York, NY, USA},
  \bibinfo{pages}{722–737}.
\newblock
\showISBNx{9798400703867}
\href{https://doi.org/10.1145/3620666.3651380}{doi:\nolinkurl{10.1145/3620666.3651380}}


\bibitem[Hu et~al\mbox{.}(2024a)]%
        {hu2024memserve}
\bibfield{author}{\bibinfo{person}{Cunchen Hu}, \bibinfo{person}{Heyang Huang},
  \bibinfo{person}{Junhao Hu}, \bibinfo{person}{Jiang Xu},
  \bibinfo{person}{Xusheng Chen}, \bibinfo{person}{Tao Xie},
  \bibinfo{person}{Chenxi Wang}, \bibinfo{person}{Sa Wang},
  \bibinfo{person}{Yungang Bao}, \bibinfo{person}{Ninghui Sun},
  {et~al\mbox{.}}} \bibinfo{year}{2024}\natexlab{a}.
\newblock \showarticletitle{Memserve: Context caching for disaggregated llm
  serving with elastic memory pool}.
\newblock \bibinfo{journal}{\emph{arXiv preprint arXiv:2406.17565}}
  (\bibinfo{year}{2024}).
\newblock


\bibitem[Hu et~al\mbox{.}(2024b)]%
        {hu2024inference}
\bibfield{author}{\bibinfo{person}{Cunchen Hu}, \bibinfo{person}{Heyang Huang},
  \bibinfo{person}{Liangliang Xu}, \bibinfo{person}{Xusheng Chen},
  \bibinfo{person}{Jiang Xu}, \bibinfo{person}{Shuang Chen},
  \bibinfo{person}{Hao Feng}, \bibinfo{person}{Chenxi Wang},
  \bibinfo{person}{Sa Wang}, \bibinfo{person}{Yungang Bao}, {et~al\mbox{.}}}
  \bibinfo{year}{2024}\natexlab{b}.
\newblock \showarticletitle{Inference without interference: Disaggregate llm
  inference for mixed downstream workloads}.
\newblock \bibinfo{journal}{\emph{arXiv preprint arXiv:2401.11181}}
  (\bibinfo{year}{2024}).
\newblock


\bibitem[Kwon et~al\mbox{.}(2023)]%
        {vllm}
\bibfield{author}{\bibinfo{person}{Woosuk Kwon}, \bibinfo{person}{Zhuohan Li},
  \bibinfo{person}{Siyuan Zhuang}, \bibinfo{person}{Ying Sheng},
  \bibinfo{person}{Lianmin Zheng}, \bibinfo{person}{Cody~Hao Yu},
  \bibinfo{person}{Joseph Gonzalez}, \bibinfo{person}{Hao Zhang}, {and}
  \bibinfo{person}{Ion Stoica}.} \bibinfo{year}{2023}\natexlab{}.
\newblock \showarticletitle{Efficient memory management for large language
  model serving with pagedattention}. In \bibinfo{booktitle}{\emph{Proceedings
  of the 29th Symposium on Operating Systems Principles}}.
  \bibinfo{pages}{611--626}.
\newblock


\bibitem[Leviathan et~al\mbox{.}(2023)]%
        {leviathan2023fast}
\bibfield{author}{\bibinfo{person}{Yaniv Leviathan}, \bibinfo{person}{Matan
  Kalman}, {and} \bibinfo{person}{Yossi Matias}.}
  \bibinfo{year}{2023}\natexlab{}.
\newblock \showarticletitle{Fast inference from transformers via speculative
  decoding}. In \bibinfo{booktitle}{\emph{International Conference on Machine
  Learning}}. PMLR, \bibinfo{pages}{19274--19286}.
\newblock


\bibitem[Li et~al\mbox{.}(2025)]%
        {li2025adaserve}
\bibfield{author}{\bibinfo{person}{Zikun Li}, \bibinfo{person}{Zhuofu Chen},
  \bibinfo{person}{Remi Delacourt}, \bibinfo{person}{Gabriele Oliaro},
  \bibinfo{person}{Zeyu Wang}, \bibinfo{person}{Qinghan Chen},
  \bibinfo{person}{Shuhuai Lin}, \bibinfo{person}{April Yang},
  \bibinfo{person}{Zhihao Zhang}, \bibinfo{person}{Zhuoming Chen},
  {et~al\mbox{.}}} \bibinfo{year}{2025}\natexlab{}.
\newblock \showarticletitle{AdaServe: SLO-Customized LLM Serving with
  Fine-Grained Speculative Decoding}.
\newblock \bibinfo{journal}{\emph{arXiv preprint arXiv:2501.12162}}
  (\bibinfo{year}{2025}).
\newblock


\bibitem[Mei et~al\mbox{.}(2025)]%
        {helix}
\bibfield{author}{\bibinfo{person}{Yixuan Mei}, \bibinfo{person}{Yonghao
  Zhuang}, \bibinfo{person}{Xupeng Miao}, \bibinfo{person}{Juncheng Yang},
  \bibinfo{person}{Zhihao Jia}, {and} \bibinfo{person}{Rashmi Vinayak}.}
  \bibinfo{year}{2025}\natexlab{}.
\newblock \showarticletitle{Helix: Serving Large Language Models over
  Heterogeneous GPUs and Network via Max-Flow} \emph{(\bibinfo{series}{ASPLOS
  '25})}. \bibinfo{publisher}{Association for Computing Machinery},
  \bibinfo{address}{New York, NY, USA}, \bibinfo{pages}{586–602}.
\newblock
\showISBNx{9798400706981}
\href{https://doi.org/10.1145/3669940.3707215}{doi:\nolinkurl{10.1145/3669940.3707215}}


\bibitem[Microsoft(2023)]%
        {microsoft_copilot}
\bibfield{author}{\bibinfo{person}{Microsoft}.}
  \bibinfo{year}{2023}\natexlab{}.
\newblock \bibinfo{title}{GitHub Copilot: AI-powered Code Assistant}.
\newblock
\urldef\tempurl%
\url{https://github.com/features/copilot}
\showURL{%
\tempurl}
\newblock
\shownote{Accessed: 2025-03-03}.


\bibitem[Moritz et~al\mbox{.}(2018)]%
        {moritz2018ray}
\bibfield{author}{\bibinfo{person}{Philipp Moritz}, \bibinfo{person}{Robert
  Nishihara}, \bibinfo{person}{Stephanie Wang}, \bibinfo{person}{Alexey
  Tumanov}, \bibinfo{person}{Richard Liaw}, \bibinfo{person}{Eric Liang},
  \bibinfo{person}{Melih Elibol}, \bibinfo{person}{Zongheng Yang},
  \bibinfo{person}{William Paul}, \bibinfo{person}{Michael~I Jordan},
  {et~al\mbox{.}}} \bibinfo{year}{2018}\natexlab{}.
\newblock \showarticletitle{Ray: A distributed framework for emerging
  $\{$AI$\}$ applications}. In \bibinfo{booktitle}{\emph{13th USENIX symposium
  on operating systems design and implementation (OSDI 18)}}.
  \bibinfo{pages}{561--577}.
\newblock


\bibitem[Muennighoff et~al\mbox{.}(2025)]%
        {muennighoff2025s1}
\bibfield{author}{\bibinfo{person}{Niklas Muennighoff}, \bibinfo{person}{Zitong
  Yang}, \bibinfo{person}{Weijia Shi}, \bibinfo{person}{Xiang~Lisa Li},
  \bibinfo{person}{Li Fei-Fei}, \bibinfo{person}{Hannaneh Hajishirzi},
  \bibinfo{person}{Luke Zettlemoyer}, \bibinfo{person}{Percy Liang},
  \bibinfo{person}{Emmanuel Cand{\`e}s}, {and} \bibinfo{person}{Tatsunori
  Hashimoto}.} \bibinfo{year}{2025}\natexlab{}.
\newblock \showarticletitle{s1: Simple test-time scaling}.
\newblock \bibinfo{journal}{\emph{arXiv preprint arXiv:2501.19393}}
  (\bibinfo{year}{2025}).
\newblock


\bibitem[Oh et~al\mbox{.}(2024)]%
        {exegpt}
\bibfield{author}{\bibinfo{person}{Hyungjun Oh}, \bibinfo{person}{Kihong Kim},
  \bibinfo{person}{Jaemin Kim}, \bibinfo{person}{Sungkyun Kim},
  \bibinfo{person}{Junyeol Lee}, \bibinfo{person}{Du-seong Chang}, {and}
  \bibinfo{person}{Jiwon Seo}.} \bibinfo{year}{2024}\natexlab{}.
\newblock \showarticletitle{ExeGPT: Constraint-Aware Resource Scheduling for
  LLM Inference}. In \bibinfo{booktitle}{\emph{Proceedings of the 29th ACM
  International Conference on Architectural Support for Programming Languages
  and Operating Systems, Volume 2}} (La Jolla, CA, USA)
  \emph{(\bibinfo{series}{ASPLOS '24})}. \bibinfo{publisher}{Association for
  Computing Machinery}, \bibinfo{address}{New York, NY, USA},
  \bibinfo{pages}{369–384}.
\newblock
\showISBNx{9798400703850}
\href{https://doi.org/10.1145/3620665.3640383}{doi:\nolinkurl{10.1145/3620665.3640383}}


\bibitem[OpenAI(2023)]%
        {openai_chatgpt}
\bibfield{author}{\bibinfo{person}{OpenAI}.} \bibinfo{year}{2023}\natexlab{}.
\newblock \bibinfo{title}{ChatGPT: Large Language Model Chatbot}.
\newblock
\urldef\tempurl%
\url{https://openai.com}
\showURL{%
\tempurl}
\newblock
\shownote{Accessed: 2025-03-03}.


\bibitem[OpenAI(2024a)]%
        {openai_batch_overview}
\bibfield{author}{\bibinfo{person}{OpenAI}.} \bibinfo{year}{2024}\natexlab{a}.
\newblock \bibinfo{booktitle}{\emph{Batch Processing Overview}}.
\newblock
\urldef\tempurl%
\url{https://platform.openai.com/docs/guides/batch/overview?lang=curl}
\showURL{%
\tempurl}
\newblock
\shownote{Accessed: 2024-12-22}.


\bibitem[OpenAI(2024b)]%
        {openai_summary}
\bibfield{author}{\bibinfo{person}{OpenAI}.} \bibinfo{year}{2024}\natexlab{b}.
\newblock \bibinfo{title}{Summarizing Long document}.
\newblock
\urldef\tempurl%
\url{https://cookbook.openai.com/examples/summarizing_long_documents}
\showURL{%
\tempurl}
\newblock
\shownote{Accessed: 2025-03-03}.


\bibitem[Patel et~al\mbox{.}(2024)]%
        {patel2024splitwise}
\bibfield{author}{\bibinfo{person}{Pratyush Patel}, \bibinfo{person}{Esha
  Choukse}, \bibinfo{person}{Chaojie Zhang}, \bibinfo{person}{Aashaka Shah},
  \bibinfo{person}{{\'I}{\~n}igo Goiri}, \bibinfo{person}{Saeed Maleki}, {and}
  \bibinfo{person}{Ricardo Bianchini}.} \bibinfo{year}{2024}\natexlab{}.
\newblock \showarticletitle{Splitwise: Efficient generative llm inference using
  phase splitting}. In \bibinfo{booktitle}{\emph{2024 ACM/IEEE 51st Annual
  International Symposium on Computer Architecture (ISCA)}}. IEEE,
  \bibinfo{pages}{118--132}.
\newblock


\bibitem[Prabhu et~al\mbox{.}(2024)]%
        {prabhu2024vattention}
\bibfield{author}{\bibinfo{person}{Ramya Prabhu}, \bibinfo{person}{Ajay Nayak},
  \bibinfo{person}{Jayashree Mohan}, \bibinfo{person}{Ramachandran Ramjee},
  {and} \bibinfo{person}{Ashish Panwar}.} \bibinfo{year}{2024}\natexlab{}.
\newblock \showarticletitle{vattention: Dynamic memory management for serving
  llms without pagedattention}.
\newblock \bibinfo{journal}{\emph{arXiv preprint arXiv:2405.04437}}
  (\bibinfo{year}{2024}).
\newblock


\bibitem[Prabhu et~al\mbox{.}(2025)]%
        {vattention}
\bibfield{author}{\bibinfo{person}{Ramya Prabhu}, \bibinfo{person}{Ajay Nayak},
  \bibinfo{person}{Jayashree Mohan}, \bibinfo{person}{Ramachandran Ramjee},
  {and} \bibinfo{person}{Ashish Panwar}.} \bibinfo{year}{2025}\natexlab{}.
\newblock \showarticletitle{vAttention: Dynamic Memory Management for Serving
  LLMs without PagedAttention}. In \bibinfo{booktitle}{\emph{Proceedings of the
  30th ACM International Conference on Architectural Support for Programming
  Languages and Operating Systems, Volume 1}} (Rotterdam, Netherlands)
  \emph{(\bibinfo{series}{ASPLOS '25})}. \bibinfo{publisher}{Association for
  Computing Machinery}, \bibinfo{address}{New York, NY, USA},
  \bibinfo{pages}{1133–1150}.
\newblock
\showISBNx{9798400706981}
\href{https://doi.org/10.1145/3669940.3707256}{doi:\nolinkurl{10.1145/3669940.3707256}}


\bibitem[Qin et~al\mbox{.}(2024)]%
        {qin2024mooncake}
\bibfield{author}{\bibinfo{person}{Ruoyu Qin}, \bibinfo{person}{Zheming Li},
  \bibinfo{person}{Weiran He}, \bibinfo{person}{Mingxing Zhang},
  \bibinfo{person}{Yongwei Wu}, \bibinfo{person}{Weimin Zheng}, {and}
  \bibinfo{person}{Xinran Xu}.} \bibinfo{year}{2024}\natexlab{}.
\newblock \showarticletitle{Mooncake: A kvcache-centric disaggregated
  architecture for llm serving}.
\newblock \bibinfo{journal}{\emph{arXiv preprint arXiv:2407.00079}}
  (\bibinfo{year}{2024}).
\newblock


\bibitem[Qin et~al\mbox{.}(2023)]%
        {qin2023toolllm}
\bibfield{author}{\bibinfo{person}{Yujia Qin}, \bibinfo{person}{Shihao Liang},
  \bibinfo{person}{Yining Ye}, \bibinfo{person}{Kunlun Zhu},
  \bibinfo{person}{Lan Yan}, \bibinfo{person}{Yaxi Lu}, \bibinfo{person}{Yankai
  Lin}, \bibinfo{person}{Xin Cong}, \bibinfo{person}{Xiangru Tang},
  \bibinfo{person}{Bill Qian}, {et~al\mbox{.}}}
  \bibinfo{year}{2023}\natexlab{}.
\newblock \showarticletitle{Toolllm: Facilitating large language models to
  master 16000+ real-world apis}.
\newblock \bibinfo{journal}{\emph{arXiv preprint arXiv:2307.16789}}
  (\bibinfo{year}{2023}).
\newblock


\bibitem[Sajal et~al\mbox{.}(2023)]%
        {sajal2023kerveros}
\bibfield{author}{\bibinfo{person}{Sultan~Mahmud Sajal}, \bibinfo{person}{Luke
  Marshall}, \bibinfo{person}{Beibin Li}, \bibinfo{person}{Shandan Zhou},
  \bibinfo{person}{Abhisek Pan}, \bibinfo{person}{Konstantina Mellou},
  \bibinfo{person}{Deepak Narayanan}, \bibinfo{person}{Timothy Zhu},
  \bibinfo{person}{David Dion}, \bibinfo{person}{Thomas Moscibroda},
  {et~al\mbox{.}}} \bibinfo{year}{2023}\natexlab{}.
\newblock \showarticletitle{Kerveros: Efficient and Scalable Cloud Admission
  Control}. In \bibinfo{booktitle}{\emph{17th USENIX Symposium on Operating
  Systems Design and Implementation (OSDI 23)}}. \bibinfo{pages}{227--245}.
\newblock


\bibitem[Sheng et~al\mbox{.}(2024)]%
        {sheng2024fairness}
\bibfield{author}{\bibinfo{person}{Ying Sheng}, \bibinfo{person}{Shiyi Cao},
  \bibinfo{person}{Dacheng Li}, \bibinfo{person}{Banghua Zhu},
  \bibinfo{person}{Zhuohan Li}, \bibinfo{person}{Danyang Zhuo},
  \bibinfo{person}{Joseph~E Gonzalez}, {and} \bibinfo{person}{Ion Stoica}.}
  \bibinfo{year}{2024}\natexlab{}.
\newblock \showarticletitle{Fairness in serving large language models}. In
  \bibinfo{booktitle}{\emph{18th USENIX Symposium on Operating Systems Design
  and Implementation (OSDI 24)}}. \bibinfo{pages}{965--988}.
\newblock


\bibitem[Song et~al\mbox{.}(2024)]%
        {powerinfer}
\bibfield{author}{\bibinfo{person}{Yixin Song}, \bibinfo{person}{Zeyu Mi},
  \bibinfo{person}{Haotong Xie}, {and} \bibinfo{person}{Haibo Chen}.}
  \bibinfo{year}{2024}\natexlab{}.
\newblock \showarticletitle{PowerInfer: Fast Large Language Model Serving with
  a Consumer-grade GPU}. In \bibinfo{booktitle}{\emph{Proceedings of the ACM
  SIGOPS 30th Symposium on Operating Systems Principles}} (Austin, TX, USA)
  \emph{(\bibinfo{series}{SOSP '24})}. \bibinfo{publisher}{Association for
  Computing Machinery}, \bibinfo{address}{New York, NY, USA},
  \bibinfo{pages}{590–606}.
\newblock
\showISBNx{9798400712517}
\href{https://doi.org/10.1145/3694715.3695964}{doi:\nolinkurl{10.1145/3694715.3695964}}


\bibitem[Stojkovic et~al\mbox{.}(2024)]%
        {stojkovic2024dynamollm}
\bibfield{author}{\bibinfo{person}{Jovan Stojkovic}, \bibinfo{person}{Chaojie
  Zhang}, \bibinfo{person}{{\'I}{\~n}igo Goiri}, \bibinfo{person}{Josep
  Torrellas}, {and} \bibinfo{person}{Esha Choukse}.}
  \bibinfo{year}{2024}\natexlab{}.
\newblock \showarticletitle{Dynamollm: Designing llm inference clusters for
  performance and energy efficiency}.
\newblock \bibinfo{journal}{\emph{arXiv preprint arXiv:2408.00741}}
  (\bibinfo{year}{2024}).
\newblock


\bibitem[Sun et~al\mbox{.}(2024)]%
        {sun2024llumnix}
\bibfield{author}{\bibinfo{person}{Biao Sun}, \bibinfo{person}{Ziming Huang},
  \bibinfo{person}{Hanyu Zhao}, \bibinfo{person}{Wencong Xiao},
  \bibinfo{person}{Xinyi Zhang}, \bibinfo{person}{Yong Li}, {and}
  \bibinfo{person}{Wei Lin}.} \bibinfo{year}{2024}\natexlab{}.
\newblock \showarticletitle{Llumnix: Dynamic scheduling for large language
  model serving}. In \bibinfo{booktitle}{\emph{18th USENIX Symposium on
  Operating Systems Design and Implementation (OSDI 24)}}.
  \bibinfo{pages}{173--191}.
\newblock


\bibitem[Williams et~al\mbox{.}(2009)]%
        {williams2009roofline}
\bibfield{author}{\bibinfo{person}{Samuel Williams}, \bibinfo{person}{Andrew
  Waterman}, {and} \bibinfo{person}{David Patterson}.}
  \bibinfo{year}{2009}\natexlab{}.
\newblock \showarticletitle{Roofline: an insightful visual performance model
  for multicore architectures}.
\newblock \bibinfo{journal}{\emph{Commun. ACM}} \bibinfo{volume}{52},
  \bibinfo{number}{4} (\bibinfo{year}{2009}), \bibinfo{pages}{65--76}.
\newblock


\bibitem[Wu et~al\mbox{.}(2024)]%
        {loongserve}
\bibfield{author}{\bibinfo{person}{Bingyang Wu}, \bibinfo{person}{Shengyu Liu},
  \bibinfo{person}{Yinmin Zhong}, \bibinfo{person}{Peng Sun},
  \bibinfo{person}{Xuanzhe Liu}, {and} \bibinfo{person}{Xin Jin}.}
  \bibinfo{year}{2024}\natexlab{}.
\newblock \showarticletitle{LoongServe: Efficiently Serving Long-Context Large
  Language Models with Elastic Sequence Parallelism}. In
  \bibinfo{booktitle}{\emph{Proceedings of the ACM SIGOPS 30th Symposium on
  Operating Systems Principles}} (Austin, TX, USA) \emph{(\bibinfo{series}{SOSP
  '24})}. \bibinfo{publisher}{Association for Computing Machinery},
  \bibinfo{address}{New York, NY, USA}, \bibinfo{pages}{640–654}.
\newblock
\showISBNx{9798400712517}
\href{https://doi.org/10.1145/3694715.3695948}{doi:\nolinkurl{10.1145/3694715.3695948}}


\bibitem[Yu et~al\mbox{.}(2022)]%
        {yu2022orca}
\bibfield{author}{\bibinfo{person}{Gyeong-In Yu}, \bibinfo{person}{Joo~Seong
  Jeong}, \bibinfo{person}{Geon-Woo Kim}, \bibinfo{person}{Soojeong Kim}, {and}
  \bibinfo{person}{Byung-Gon Chun}.} \bibinfo{year}{2022}\natexlab{}.
\newblock \showarticletitle{Orca: A distributed serving system for
  $\{$Transformer-Based$\}$ generative models}. In
  \bibinfo{booktitle}{\emph{16th USENIX Symposium on Operating Systems Design
  and Implementation (OSDI 22)}}. \bibinfo{pages}{521--538}.
\newblock


\bibitem[Zeng et~al\mbox{.}(2025)]%
        {medusa}
\bibfield{author}{\bibinfo{person}{Shaoxun Zeng}, \bibinfo{person}{Minhui Xie},
  \bibinfo{person}{Shiwei Gao}, \bibinfo{person}{Youmin Chen}, {and}
  \bibinfo{person}{Youyou Lu}.} \bibinfo{year}{2025}\natexlab{}.
\newblock \showarticletitle{Medusa: Accelerating Serverless LLM Inference with
  Materialization} \emph{(\bibinfo{series}{ASPLOS '25})}.
  \bibinfo{publisher}{Association for Computing Machinery},
  \bibinfo{address}{New York, NY, USA}, \bibinfo{pages}{653–668}.
\newblock
\showISBNx{9798400706981}
\href{https://doi.org/10.1145/3669940.3707285}{doi:\nolinkurl{10.1145/3669940.3707285}}


\bibitem[Zhang et~al\mbox{.}(2022)]%
        {zhang2022opt}
\bibfield{author}{\bibinfo{person}{Susan Zhang}, \bibinfo{person}{Stephen
  Roller}, \bibinfo{person}{Naman Goyal}, \bibinfo{person}{Mikel Artetxe},
  \bibinfo{person}{Moya Chen}, \bibinfo{person}{Shuohui Chen},
  \bibinfo{person}{Christopher Dewan}, \bibinfo{person}{Mona Diab},
  \bibinfo{person}{Xian Li}, \bibinfo{person}{Xi~Victoria Lin},
  \bibinfo{person}{Todor Mihaylov}, \bibinfo{person}{Myle Ott},
  \bibinfo{person}{Sam Shleifer}, \bibinfo{person}{Kurt Shuster},
  \bibinfo{person}{Daniel Simig}, \bibinfo{person}{Punit~Singh Koura},
  \bibinfo{person}{Anjali Sridhar}, \bibinfo{person}{Tianlu Wang}, {and}
  \bibinfo{person}{Luke Zettlemoyer}.} \bibinfo{year}{2022}\natexlab{}.
\newblock \bibinfo{title}{OPT: Open Pre-trained Transformer Language Models}.
\newblock
\showeprint[arxiv]{2205.01068}~[cs.CL]


\bibitem[Zheng et~al\mbox{.}(2024)]%
        {zheng2024sglang}
\bibfield{author}{\bibinfo{person}{Lianmin Zheng}, \bibinfo{person}{Liangsheng
  Yin}, \bibinfo{person}{Zhiqiang Xie}, \bibinfo{person}{Chuyue~Livia Sun},
  \bibinfo{person}{Jeff Huang}, \bibinfo{person}{Cody~Hao Yu},
  \bibinfo{person}{Shiyi Cao}, \bibinfo{person}{Christos Kozyrakis},
  \bibinfo{person}{Ion Stoica}, \bibinfo{person}{Joseph~E Gonzalez},
  {et~al\mbox{.}}} \bibinfo{year}{2024}\natexlab{}.
\newblock \showarticletitle{Sglang: Efficient execution of structured language
  model programs}.
\newblock \bibinfo{journal}{\emph{Advances in Neural Information Processing
  Systems}}  \bibinfo{volume}{37} (\bibinfo{year}{2024}),
  \bibinfo{pages}{62557--62583}.
\newblock


\bibitem[Zhong et~al\mbox{.}(2024)]%
        {zhong2024distserve}
\bibfield{author}{\bibinfo{person}{Yinmin Zhong}, \bibinfo{person}{Shengyu
  Liu}, \bibinfo{person}{Junda Chen}, \bibinfo{person}{Jianbo Hu},
  \bibinfo{person}{Yibo Zhu}, \bibinfo{person}{Xuanzhe Liu},
  \bibinfo{person}{Xin Jin}, {and} \bibinfo{person}{Hao Zhang}.}
  \bibinfo{year}{2024}\natexlab{}.
\newblock \showarticletitle{Distserve: Disaggregating prefill and decoding for
  goodput-optimized large language model serving}.
\newblock \bibinfo{journal}{\emph{arXiv preprint arXiv:2401.09670}}
  (\bibinfo{year}{2024}).
\newblock


\end{thebibliography}

\appendix
\newpage

\section{Detailed Analysis into the disaggregated Scheduling.}\label{apdx:disagg-sch}

In disaggreagated scheduling~\cite{zhong2024distserve,patel2024splitwise}, different stages are separated onto different (group of) devices. In this way, one can specialize for every stage's SLO by adjusting the hardware configurations~\cite{zhong2024distserve}, and balance workloads across stages by customizing adjusting the devices allocated to every stages.

The flexibility of disaggregated scheduling is hindered by the fact that the optimal hardware configuration depends on the load across stages, making it a hard fit for our setup where every stage has dynamic load characteristics. Shown in Fig.~\ref{fig:profile-disagg}, decode-heavy workloads like coding requires more decoding device allocation (1:2), whereas prefill-heavy workloads like chatting requires more prefill device allocation (2:1). As a result, not a single device allocation strategy works across different requests, resulting in poor performance when the requests have mixed load-patterns.

For the simplicity of analysis, we assumes an ideal setting where the prefill devices is operating under the maximum throughput and the system scales perfectly with the number of devices. 
Suppose the maximum goodput is $g$, then for the prefill, the token budgets generated by all prefill devices per time unit must match the new tokens need to be processed
\begin{align*}
    g \times E_r[InputLength_r]  \le MaxTokenTpt \times n_{prefill}
\end{align*}, where $n_{prefill}$ denotes the number of prefill devices.  

For decode, every request in the system contributes to one token budget in the next batch. Therefore, the time to execute this batch should below the TPOT requirement. Particularly,
\begin{align*}
    BatchTime(\#ReqInSystem) &\le TPOT\\
    \#ReqInSystem \times n_{decode} &= g \times E_r[TPOT \times OutputLength_r]
\end{align*}
By approximating the $BatchTime$ function using
\begin{align*}
    BatchTime(n) = \frac{n}{MaxTokenTpt} + Overhead
\end{align*}
We have
\begin{align*}
    g &\le (1-\frac{Overhead}{TPOT}) \frac{MaxTokenTpt\times n_{decode}}{E_r[OutputLength_r]} 
\end{align*}

Hence, $g$ is bounded by 
\begin{align*}
\scriptsize
g \le MaxTokenTpt\times \frac{min(\frac{ n_{prefill}} {E_r[I_r]}, 
(\frac{TPOT-Overhead}{TPOT}) \frac{ n_{decode}}{E_r[O_r]})}{ n_{decode} + n_{prefill}}
\end{align*}
By optimizing the device allocation $n_{prefill}$ and $n_{decode}$, we have 
\begin{align}
    g^* = \frac{MaxTokenTpt}{\frac{TPOT}{TPOT - Overhead}E_r[O_r] + E_r[I_r]}\label{eqn: max goodput}
\end{align}
Under the condition




Note that the balanced ratio (Eqn.~\ref{eqn: optimal proportion}) between the prefill devices and the decoding devices are influenced by two factors (i) the SLO requirement, and (ii) input/output lengths of requests. Consequently, service providers must reconfigure their systems to adapt to evolving service requirements.


The dependency on SLO and token length patterns highlights the foundamental difficulty of disaggregated scheduling to support multi-tier services.  Specifically, when requests exhibit drastically different input length/output length patterns as well as SLOs, no fixed configuration can achieve balanced device allocation, leading to inefficient device utilization. 

\section{More on LLM Serving}

Unlike traditional applications such as web search and online data mining, where strict run-time budgets constrain request scheduling, LLM serving involves longer request lifespans and coarser request arrivals. This allows for more advanced and comprehensive scheduling algorithms. For instance, statistics from common serving frameworks (Tab.~\ref{tab:typical}) indicate that a scheduler invoked every second typically handles around 0.35 to 100 requests per second, providing ample opportunity for well-optimized scheduling decisions.

\begin{table}[]
\caption{Common Statistics in a LLM serving framework. The request rate data is obtained from \cite{zhong2024distserve,sarathi}.}
\label{tab:typical}
\resizebox{\linewidth}{!}{
\begin{tabular}{@{}cccc@{}}
\toprule
Request Rate & Lifespan & Prefill Lifespan & Decode Lifespan\\ \midrule
0.5-10 Req/s    & 0.7-10s    & 0.1-1s                & 0.5-8s                           \\ \bottomrule
\end{tabular}
}
\end{table}

\section{Time Complexity Analysis}\label{apx:time-complexity}
When optimizing for request throughput, i.e. the number of request accepted, the DP state in Eqn.~\ref{eqn:dp} directly encodes the objective function. 
Under this setup, we can refactor the DP to a $pd[i,m,n]$ that calculates the maximal prefill token budgets available by request $i$'s prefill deadline under $m$ memory units while accepting $n$ requests.
\begin{align}
&pb[i, m, n] \nonumber\\
&= \max_{\substack{j,\\ pDDL_j < pDDL_i, \\ \Delta pb \ge 0,\  m \ge m_i}} \left\{ pb[j, m - m_i, n-1] - p_i + \Delta pb \right\}\label{eqn:dp}.
\end{align}
Therefore, the time complexity is reduced to $O(N\times M \times N_{new}^L)$

\begin{algorithm}
\caption{BatchForward At Time Step t}
\label{apx:batch-forward}
\KwIn{$r_{1...m}: \text{prefill requests}$, $R_{1...n}: \text{decode request}$, $p^{t}_{1..m}: \text{prefill token per request}$, $s^{t}_{1..n}: \text{decode token per request}$}

\KwState{$C^{request, model}_{b:e}: \text{KV Caches for }\ request\ \text{on}\ model\ \text{for tokens from } b\ \text{to}\ e.$,
$x^{request, model}_{b:e}: \text{input  tokens for }\ request\ \text{on}\ model\ \text{for tokens from } b\ \text{to}\ e.$
$y^{request, model}_{b:e}: \text{predicted tokens for }\ request\ \text{on}\ model\ \text{for tokens from } b\ \text{to}\ e.$}

\KwOut{Results of the batch forward process for drafter and base models}

Perform prefill on the drafter model\;
$[y^{R, drafter}_{m_R:m_R + p^t_R}]_R, [C^{R, drafter}_{m_R:m_R + p^t_R}]_R \gets BatchFWD_{drafter}([x^{R, drafter}_{m_R:m_R + p^t_R}]_R, [C^{R, drafter}_{1:m_R}]_R$)\;\label{exec:drafter-prefill}

Perform iterative decoding on the drafter model to calculate speculations\;
\For{$k = 1$ \KwTo $\max_r(s^t_r)$}{
    $[y^{r, drafter}_{n_r + k}]_r, [C^{r, drafter}_{n_r + k}]_r \gets$ $BatchFWD_{drafter}([x^{r, drafter}_{1:n_r + k}]_r, [C^{r, drafter}_{1:n_r + k}]_r)$\label{exec:drafter-decode}
}

Perform batched prefill and verification on the base model\;
$[y^{r, base}_{n_r + 1:n_r + s^t_r}]_r + [y^{R, base}_{m_R:m_R + p^t_R}]_R,[C^{r, base}_{n_r + 1:n_r + s^t_r}]_r + [C^{R, base}_{m_R:m_R + p^t_R}]_R \gets BatchFWD_{base}(
    [x^{r, drafter}_{n_r + 1:n_r + s^t_r}]_r + [x^{R, base}_{m_R:m_R + p^t_R}]_R,
    [C^{r, base}_{1:n_r}]_r + [C^{R, base}_{1:m_R}]_R$)\;\label{exec:base-forward}

Verify the correctness of the speculation\;
$[y^{r, base}_{n_r + 1:n_r + a^r_t}]_r \gets$ BatchVerify($[y^{r, drafter}_{n_r + 1:n_r + s^r_t}]_r, [y^{r, base}_{n_r + 1:n_r + s^r_t}]_r$)\;\label{exec:verify}

\end{algorithm}

\section{SLO-adaptive Speculative Decoding}~\label{apx:ada-spec}

We now explain how to solve the following problem:
\begin{align*}
\max_{\substack{sl_{1:L}}} prefillTpt &:= \frac{\text{PrefillBgtPerBatch}}{\text{Batch Time}}  \\
\text{PrefillBgtPerBatch} &= Time2BS(T(sl_{1:L}), sl_{1:L}) - \sum_i n_i sl_i \\
\text{Batch Time} &= T(sl_{1:L}) = \min_{l \in 1, 2, ..., L}(TPOT_l \cdot Acc(sl_l)) 
\end{align*}

First, suppose the speculation accuracy is $\alpha$, $Acc(sl_l) = \frac{1-\alpha^l}{1-\alpha}$. Suppose $l^* = arg\min_{l \in 1, 2, ..., L}(TPOT_l \cdot Acc(sl_l))$, it is easy to other $l'$s satisfy $sl_{l'} = arg\min TPOT_{l'} \cdot Acc(sl_{l'}) > TPOT_{l} \cdot Acc(sl_{l})$, which has a close-formed solution. Therefore, we just need to enumerate $l$ that minimizes $\min_{l \in 1, 2, ..., L}(TPOT_l \cdot Acc(sl_l))$ and find one that generates largest prefill throughput. This takes constant times in practice, given the maximum speculation decode lengths is belowe 10 and there is 2-3 different decode SLOs.









\end{document}